\documentclass[manuscript]{aastex}

\newcommand{\kms}{km~s$^{-1}$}
\newcommand{\gal}{$\alpha$}
\newcommand{\gb}{$\beta$}
\newcommand{\gla}{$\lambda$}
\newcommand{\etal}{et al.}
\newcommand{\chandra}{{\it Chandra}} 

\slugcomment{ApJ, accepted February 24, 2012}

\shorttitle{TW Hya}
\shortauthors{Dupree et al.}

\begin{document}


\title{TW Hya: Spectral Variability, X-Rays, and Accretion Diagnostics}


\author{A. K. Dupree,
N. S. Brickhouse,  S. R. Cranmer,  G. J. M. Luna\altaffilmark{1}, E. E. 
Schneider\altaffilmark{2}}
\affil{Harvard-Smithsonian Center for Astrophysics, Cambridge, MA
  02138, USA}

\author{M. S. Bessell}
\affil{Australian National Observatory, Mount Stromlo Observatory,
ACT, Australia}

\author{A. Bonanos}
\affil{Institute of Astronomy and Astrophysics, National Observatory
  of Athens, 
15236 Athens, Greece}

\author{L. A. Crause}
\affil{South African Astronomical Observatory, P.O. Box 9, Observatory
  7935, South Africa}

\author{W. A. Lawson}
\affil{University of New South Wales, Canberra, ACT 2600, Australia}

\author{S. V. Mallik}
\affil{Indian Institute of Astrophysics, Bangalore 560034, India}

\author{S. C. Schuler}
\affil{National Optical Astronomy Observatory, Tucson, AZ, 85719, USA}

\altaffiltext{1}{Present address: Instituto de Astronomia y Fisica del
  Espacio, (IAFE), Buenos Aires, Argentina}

\altaffiltext{2}{Present address: Astronomy Department, University of
  Arizona,
Tucson, AZ}

\begin{abstract}
The nearest accreting T Tauri star, TW Hya was intensively and
continuously observed  over $\sim$17 days with spectroscopic and
photometric measurements from 4 continents simultaneous with  
a long segmented exposure
using the \chandra\ satellite. Contemporaneous
optical photometry from WASP-S indicates  a 4.74~d period was 
present during this time. Absence of a similar
periodicity  in the H-\gal\ flux and the total X-ray
flux which are dominated by accretion processes and the
stellar corona respectively points to a different source of 
photometric variations. The H-\gal\ emission line appears intrinsically 
broad and symmetric, and both the  profile and its  variability suggest 
an origin in  the post-shock cooling region.     An accretion event, signaled by
soft X-rays, is traced  spectroscopically for the first time through the optical
emission line profiles.  After the accretion event, 
downflowing turbulent material observed in  the H-\gal\ and H-\gb\ lines is followed by
\ion{He}{1} (\gla5876) broadening near the photosphere. Optical veiling
resulting from the heated photosphere increases with a delay of $\sim$ 2 hours
after the X-ray accretion event.  The response of the stellar
coronal emission to an increase in the veiling follows $\sim$2.4 hours later,
giving direct evidence  that the stellar corona is
heated in part by  accretion. Subsequently, the stellar wind becomes
re-established. We suggest a model that incorporates
the dynamics of this sequential series of events: an accretion shock, a cooling
downflow in a supersonically turbulent region, followed by 
photospheric and later, coronal heating.  This model naturally
explains the presence of broad optical and  ultraviolet lines, and affects the
mass accretion rates determined from emission line profiles.

\end{abstract}


\section{Introduction}
The young dwarf star TW Hya\footnote{V= 11.1; RA=11$^{hr}$01$^m$52$^s$;
  DEC=$-$34$^{\circ}$42'17.03'' (FK5 2000.)} (CD $-$34 7151; TWA 1; HIP 53911) is
arguably the closest accreting T Tauri star (Wichmann \etal\ 1998), and as such offers an
opportunity to study the accretion process and properties of the stellar
atmosphere and wind. Of particular interest is the fact that the 
accretion disk surrounding TW Hya is oriented almost `face-on' and the
rotation axis of the star has a low inclination (Krist \etal\ 2000;
Qi \etal\ 2004).  Such an orientation places 
the polar regions of the star in full view  where accretion
of material from a surrounding circumstellar disk 
is thought to occur.  Moreover,  it is likely that open magnetic 
field structures exist at the poles, where a 
stellar wind may arise. TW Hya presents a bright target with rich 
scientific potential to unravel both 
the accretion process and its relation to coronal heating  
and accretion-driven winds from young stars.

TW Hya has been well-studied  and monitored with  optical and ultraviolet spectra
(Muzerolle \etal\ 2000, Alencar \&
Batalha 2002; Ardila \etal\ 2002; Batalha \etal\ 2002; Dupree \etal\
2005; Curran \etal\ 2011), exhibiting variations  
in  emission line profiles that are attributed principally 
to  accretion and the presence of a stellar wind.

An X-ray spectrum of TW Hya obtained with the \chandra\  Observatory
in 2000 first indicated very high densities that
Kastner et al. (2002) attributed to the shock arising from material 
accreting  from the surrounding circumstellar disk in
magnetically-channeled
flows.  A long (500ks) pointing by the
\chandra\ satellite with the HETG grating   
confirmed those densities, and differentiated the emission lines  
from the accretion shock and the stellar corona (Brickhouse \etal\ 2010). Importantly,
the X-ray diagnostics revealed a large coronal volume   
resulting from heating by the shocked plasma.
The \chandra\  spectrum,
spectral diagnostics, and model from these definitive observations
have been discussed elsewhere (Brickhouse \etal\ 2010). Here,
we report on simultaneous and contemporaneous optical and
infrared, photometry and spectroscopy.
In particular, the behavior of X-rays
and the strength and profiles of the hydrogen and helium emission 
lines   are investigated because  simultaneous measures can elucidate 
their relationships.

These photometric and spectroscopic observations were obtained
at a variety of sites  that span times of the   
\chandra\ pointings. 
TW Hya is located at a declination of $-$35$^\circ$ making it
accessible from the northern hemisphere several hours per night during
transit, in addition to easy access from the
southern hemisphere.
Characteristics of the individual observations follow below.  Table 1
contains a summary of the \chandra\ 
observations; Table 2 summarizes the optical campaign.
Figure 1 displays the observations at  various observatories in
conjunction with the 4 segments of the \chandra\ pointings.

Photometry is addressed first (Section 2), in order to establish a
photometric period for the star contemporaneously with the
\chandra\  observations and the spectroscopy.  In Section 3, the 
X-ray measurements,  deconstructed into a coronal
and accretion component, are  compared with the optical photometry to
isolate accretion and flaring events. Section 4 summarizes the 
optical and infrared spectroscopy efforts.
Subsequent sections address the variation of the 
H-$\alpha$ and H-$\beta$ fluxes from the monitoring program (Section 5)
and an intensive high 
resolution optical study of the H-$\alpha$, H-$\beta$ and \ion{He}{1}  profiles
over 3 successive nights.  The following sections detail
the optical spectroscopic response to the X-ray accretion events and
a stellar X-ray flare (Sections 6 and 7).  The wind, as detected in
the near-infrared line of \ion{He}{1} is compared to 
previous measures and to the near-simultaneous optical spectrum 
in Section 8. Section 9 contains the veiling parameters 
and explores their relation to the X-ray
flux. 
A model incorporating our results is put forth in Section 10, and
comparison with previous results is made in Section 11.
Conclusions can be found in  Section 12.

\section{Photometric Observations: Super WASP-S, SAAO, and ASAS}

Photometry from the Super WASP-South 
program (Butters \etal\ 2010) 
spanned the \chandra\  
observations, and a period can be
sought from those data.  The measurements shown
in the top panel of Figure 2, contained a flaring episode
on JD 2454155 (Day 55)\footnote{ 
Here the nomenclature `Day 55' represents the Julian Date minus an
offset: JD$-$2454100.}. A period search 
on the non-flaring data (see Figure 2, {\it lower panel}) was made using the 
algorithm of Horne \&\ Baliunas (1986)
and Scargle (1982), which accommodates unequally spaced data.  
A clear peak occurs in the periodogram corresponding to 4.744 days during
this span of 19.2 days from JD 2454148.33 to JD 2454167.48.  The 
False Event probability assigned to this period is less than 10$^{-5}$.  A long
history of searches for a photometric  period of  TW Hya can be found
in the literature (cf. Rucinski \etal\ 2008), 
which reveals values ranging  from 1.3 to $\sim$6
days. Periods around 4.74 d have been seen in B-band veiling
(Batalha \etal\ 2002), and in some line intensity data (Alencar \& Batalha
2002) and suggested by B-band photometry (Lawson \& Crause 2005). 
While the photometric variations might 
arise from the presence of accretion hot spots or rings 
on the stellar surface, or, alternately as suggested by Siwak \etal\
(2011), from optically thick plasma condensations
in the inner accretion disk,  a clear
periodicity is  confounded by flaring episodes.  No 
long-term periodicity could be identified in  {\it Hipparcos} photometry 
(Kastner et al. 1999), although Koen and Eyer (2002) identified a 2.88
day period, which corresponds to the V-band photometry
of Lawson \& Crause (2005). Continuous observations over a short time span appear
to yield a reliable value of the period, but these values
can differ one from another, and it is not clear what phenomenon or
combination of effects is represented by a `period'.
The study by Rucinski \etal\ (2008)
analyzing photometry from the Microvariability \& Oscillation of 
STars (MOST) satellite during a continuous 11~d observation in 2007 yielded a  
periodicity of 3.7~d and the amplitude spectrum contained
significant  real components with  periods longer than 0.5~d.   The MOST observation 
followed the $\sim$19~d WASP-S segments considered in this paper by $\sim$7~d; however,  
the periodogram from the WASP-S data shows no peak at the period found
from MOST. 
Rucinski et al. (2008) noted that in 2008, the oscillations appeared 
to move towards shorter periods, from  5--6 days to 3 days over a time
interval of 10 days.  A systematic period-shortening on
time scales of weeks is confirmed in the 2009 MOST photometry as well 
(Siwak et al. 2011).  Thus, the value found here is not inconsistent
with the MOST measurements. 

Precision velocities from spectroscopic measurements made in 
February-March 2008 suggested a period of 
3.57 days (Hu\'elamo \etal\ 2008) which was
attributed to a cool spot at high latitudes.  Rucinski \etal\ (2008)
did not detect this period from MOST photometry made at the
same time. The WASP photometry (Figure 2) also does  not indicate a period
of 3.57 d (0.280 d$^{-1}$).  It seems possible that the photometry
traces variability related to the accretion process in some way. 
And so, we assume the period of 4.74~d as 
the fiducial period for comparison
here because  it was  derived from WASP-S measurements spanning the time of
the time of these X-ray and optical measurements.  When 
the photometry of TW Hya is phased to the
4.74~d period (see Figure 3), variability on shorter time
scales is also present.  Such variability is typical of T Tauri stars.

Differential V-band photometry at several epochs was obtained at the Sutherland
field station of the South African Astronomical Observatory (SAAO) 
with a 1-m telescope and a 
1~k~$\times$~1~k SITe charge-coupled device (CCD) at SAAO between 
2007 February 21-27 and 2007 May 2-8. The SITe CCD has a field of view of
26 arcmin$^2$ 
at the f/16 Cassegrain focus of the 1-m telescope. Of the 30-40 other 
stars recorded per field, several bright stars were used as local 
differential standards. The observations were made  
differentially as this allowed useful data to be obtained under 
conditions that were not always photometric. The methods used to 
produce these photometric observations are 
discussed in detail by Lawson \etal\ (2001) and Lawson \& Crause (2005).
The SAAO results are shown in Figure 4 ({\it panel C}) along with 
the WASP-S photometry ({\it panel B}), and the \chandra\ first order
flux ({\it panel A}, also see discussion
in the following section).  The flaring episode during Day 55 found in
the WASP-S data  
can also be seen in the SAAO photometry.
Additionally, 
the  X-ray flare event near Day 58.5 in Figure 4 appears 
related to an optical flare about 20 minutes later  
during which the SAAO photometry shows a brightening of $\sim$0.1
magnitude. Outside of the flaring episodes, the SAAO photometry
is consistent with the 4.74~d period.

Photometric data in V-band is available from the All Sky Automated
Survey (ASAS) of the University of Warsaw (Pojmanski 2002) located
at the Las Campanas Observatory of the Carnegie Institution of
Washington.  These fully-automated photometric monitoring measurements 
are publicly available from {\it http://www.astrouw.edu.pl/asas/}.
They are shown in the panel D of Figure 4 where the variation
is in harmony with the 4.74 day period identified in the WASP-S data,
although the cadence of ASAS ranges from 2 to 3 days.
Of particular interest is the high point near Day 57.6 which appears
to signal a brightening event contemporaneous with  the increase in
the \chandra\ accretion line flux (see Section 6).

Thus, photometric variability results from a variety of physical
processes in these accreting systems, including flaring
episodes superposed on a light modulation of varying periods.
These results illustrate how challenging it is to seek a photometric
period for TW Hya.  We adopt the photometric period of 4.744 days as
the period of reference for these observations.  None of our
results depend on the observed period.

\section{\chandra\ X-ray Observations}

HETG spectra  were obtained during 4 separate \chandra\ pointings
comprising a Large Observing Project that totaled 500ks and  
spanned the interval between 2007 February 15 and 2007 March 3 (Table 1).
These spectra, their plasma diagnostics, and the model
implied by the X-ray observations  are discussed in 
detail in Brickhouse \etal\ (2010).   These authors emphasized 
the importance of recognizing that the \chandra\ first order X-ray flux includes 
both emission lines and continua arising in different parts of the
TW~Hya system.  The {\it total} first order flux (spanning 2--27.5\AA) is  dominated by 
continuum emission produced by the stellar coronal plasma 
at $\sim$10$^7$K. In addition, the spectrum  allows
separation of emission lines according to their origin.  The high
temperature
(1.2$\times$10$^7$K) emission lines of \ion{Mg}{12}, \ion{Si}{13}, and
\ion{Si}{14} arise in the corona. The lower temperature
(2.5$\times$10$^6$K) emission lines (\ion{N}{7}, \ion{O}{8},
\ion{Ne}{9}, \ion{Fe}{17}, and \ion{Mg}{11}) form in the accretion
shock as defined in the model of Brickhouse \etal\ (2010).  These will
be referred to as X-ray accretion lines. The  newly
identified large post-shock coronal region produces the \ion{O}{7} emission.
For  comparisons with the optical spectra and photometry in the
following sections, 
we use either the total first order flux
(representing  the stellar corona) or a selection of emission lines 
originating either in the accretion shock  or the corona as defined above.

No periodicity in the  first-order X-ray flux can be 
found from analysis of the totality of observations.  Comparison of the
optical photometric period of 4.74~d  to the X-ray flux (Figure 4) shows
modulation during the third \chandra\ segment on Day 60-62 in agreement with 
photometry.  However the preceding and 
following segments do not follow the  photometric curve and the agreement 
on Day 60--62 may be coincidental. 
The first \chandra\  segment appears anomalously low  when compared
to the photometric period,  unless
there is a deviation from the average coronal flux during this 
segment. 

A lack of correlation may not be surprising since the optical light
giving the photometric period consists of variable continuum emission
from the star itself, the hot accretion regions (spots/rings),  
the activity from plages and active
regions, flaring episodes, and possibly a contribution from the
accretion disk. In any case,  {\it the total coronal emission is  not 
correlated with the optical photometric period during these 
observations.} 
This lack of X-ray periodicity is consistent with the results
of Stassun \etal\ (2006; 2007) from the young Orion Nebula Cluster where
95\% of the large COUP sample exhibited no link between X-ray
variability and optical photometry.

The X-ray spectrum of TW Hya offers additional insight (Brickhouse
\etal\ 2010). Figure 5 includes the \chandra\ 
total flux  ({\it top panel}),  
the emission line flux from  the corona 
({\it middle panel}) and
the accretion shock ({\it lower panel}). The corona is principally responsible 
for the flare during Day 58, as well as the slow increase in flux
during Day 61.  Enhancement episodes observed 
in the accretion emission lines occurred near Day 48 and 57 (marked by
arrows in Figure 5).  Separating out these components, 
we find that accretion may contribute slightly to the total  
X-ray variability.  Accretion appears to cause
the brightening of the total flux during the first \chandra\ interval
(Day 48.2) as well as  at the start of the
second \chandra\ segment (Day 57.5).  

We find that {\it variation in the total X-ray flux 
arises from both accretion and coronal
variability and neither the accretion nor coronal components 
exhibits the periodicity of the optical
photometry.}  Flaring episodes in the corona can be identified
with some brightenings found in the optical photometry.  In the
following sections, the X-ray  variations are related to the optical
line profiles.

We emphasize that the observed X-ray accretion line flux represents a
complicated mix of temperature, column density in the accreting
material, and the electron density.  The accretion rate could
increase, due to  enhancement of the accretion column density, which  for a fixed 
filling factor, absorbs and hence reduces the observed soft X-ray line flux.
Or, the accretion rate could also increase creating a larger
filling factor, accompanied by a decrease in absorbing  column density, and an
increased soft X-ray line flux. In Section 6, we explore the optical
spectroscopic response to the observed increased accretion line flux.

\section{Optical and Infrared Spectroscopic Observations}

Spectra were obtained
from 7 observatories spanning the \chandra\ pointings 
in order to follow the variation in the strength
and appearance of the emission profile, to search for the relationship
between the accretion process, the stellar atmosphere and the stellar wind, and
to relate any line variation to the X-ray flux from the corona and the accretion shock. We
summarize the telescopes and instruments in the following
sections (also Table 2), and then present an examination of line
profiles in the sections following.

\subsection{Telescopio Nationale Galileo (TNG) - Canary Islands}
Telescope time  on the 3.58--m Telescopio Nationale Galileo (TNG) was kindly 
made available through the Director's Discretionary program to
coincide
with the first CHANDRA segment during the nights
of 2007 February 17-18. Unfortunately,  one night was clouded out.
The SARG, a high efficiency echelle spectrograph with a mosaic of 
two,  2$\times$4K CCD detectors  was
binned 2$\times$2 pixels.  Aperture 2 was selected, with a slit size
of 0.8$\times$5.3 arcsec giving 
a spectral resolution of $\sim$57,000.    Luca
DeFabrizio who acquired the data in service mode kindly   
reduced the spectra using  modified IRAF\footnote{Available at {\it
    http://iraf.noao.edu/}} software.

\subsection{Vainu Bappu Observatory (VBO)- India}
The 2.3--m telescope of Vainu Bappu Observatory in Kavalur, India
made observations of TW Hya for 5 nights from 27 Feb. to 3 Mar. 2007.
Exposure times of 2700 s were used for all spectra. 
The fibre-fed echelle spectrograph provided a 
resolution of 28,000.  No sky subtraction  was applied. The spectra
were reduced with IRAF, and continuum normalization excluded the
H-$\alpha$ emission.   A preliminary report of
these observations is given in Mallik et al. (2010).

\subsection{South Africa Astronomical Observatory (SAAO) - South Africa}
Medium-resolution spectroscopy of TW Hya was obtained with the 1.9-m telescope at 
the Sutherland field station of the South African Astronomical
Observatory (SAAO)  between 2007 February 21-28. 
The Cassegrain spectrograph was used with grating `5' that 
has a 2-pixel resolution at H-$\alpha$ of 1 \AA\ ($\lambda/\Delta \lambda$ $\sim$6000). Exposure times 
ranged between 90-1800~s in order not to saturate the 
peak of the H-$\alpha$ emission line profile, or to 
obtain higher signal-to-noise ratio measurements 
of the H-$\alpha$ line wings and surrounding continuum.
The spectra were reduced with IRAF$^1$. The signal to noise 
ratio ranged from 50 to 100 in the continuum.

\subsection{Mt. Stromlo and Siding Spring Observatory (MSSO) - Australia}

Spectroscopy of TW Hya was obtained with the 2.3-m telescope and 
echelle spectrograph at Siding Spring Observatory (SSO) 
between 2007 March 1-6. The SSO spectra have a 2-pixel resolution of 
0.4\AA\ at H-\gal ($\lambda/\Delta\lambda \sim$16,400). Exposure 
times of 600-1200 s yielded spectra with 
SNR of 50 to 125 in the nearby continuum.

\subsection{Las Campanas Observatory (Magellan:Clay) - Chile}
The MIKE double-echelle spectrograph (Bernstein \etal\ 2003) 
mounted on the 6.5--m Magellan/CLAY telescope at Las Campanas
Observatory, Chile,  was
dedicated for 3 consecutive nights (26--28 February 2007) to obtain several
hundred echelle spectra spanning the optical region. An additional spectrum
was obtained on 1 March 2007.  A slit of
0.75$\times$5~arcs was used yielding a 2-pixel resolution of $\Delta
\lambda/\lambda \sim$26000 on
the blue side ($\lambda\lambda$3350--5000) and  $\sim$36,000 on the 
red side ($\lambda\lambda$4900--9300).  Spectra were taken 
with short exposures ranging from 45--90 s 
(to avoid saturation of the H-$\alpha$ 
emission) mixed with  longer exposures (360 to 600 s)  to detect weaker features.  The
IDL pipeline developed by S. Burles, R. Bernstein, and
J. S. Prochaska\footnote{See
  http://www.lco.cl/telescope-information/magellan/instruments/mike/idl-tools} was used to
extract the spectra, and IRAF software for analysis.

\subsection{Pico dos Dias Observatory - Brazil}
Optical spectra were obtained at Observatorio  Pico
dos Dias, Brazil, during the nights from 26 February through 1 March 2007.
The Coud\'e spectrograph attached to the 1.6m telescope was used with  the WI098
CCD and the 600 $\ell$/mm grating, resulting in a resolution of 
$\lambda/\Delta\lambda \sim$ 13,000  at 6,500 \AA. The
average exposure time was 1800 s.

\subsection{Gemini South - Chile}
The PHOENIX IR spectrograph on the 8--m  Gemini-South telescope was used to record
the spectrum of the He I 10830\AA\ line on 1 March 2007 
with a 4 pixel slit giving a resolution of $\lambda/\Delta\lambda \sim$50,000. Two exposures
of 900~s each 
were nodded by 4~arcsec on the CCD. 
Flat fielding of the exposures was performed with {\it IRAF} routines.
The known dispersion  of PHOENIX at this grating position was
applied to the spectra   and they were  aligned to   the 
\ion{Mg}{1} line at 10811.11\AA.

\section{The H-\gal\ and H-\gb\ Lines}

The H-\gal\ line emission dominates the optical
spectrum of  TW~Hya and its enhanced width is 
generally believed to signal the presence of 
accretion in young stars (Bertout 1989; 
Hartmann \etal\ 1994; Alencar \& Batalha 2002). 
In this section, night-to-night variability in the line
strength is first addressed from the sequence of observations
spanning 19 d, followed by a discussion of the line profiles 
of H-\gal\ and H-\gb.

\subsection{Variability of H-\gal\ Equivalent Widths}
All H-\gal\ spectra were reduced to a heliocentric
scale and  normalized to the neighboring
continuum using the {\it IRAF} tool {\it continuum} generally
with a 3 or 5th order Legendre function. 
The region  near the H-\gal\ emission line was omitted from  the
continuum fit. Averages of the profiles
are shown in Figure 6 where high
emission during four  consecutive days is  followed by lowered emission.   
Equivalent widths were obtained by integrating the continuum normalized spectra over
a 25\AA\ region from 6550 to 6575\AA.  
Comparison between the X-ray accretion line fluxes, the
H-$\alpha$ equivalent widths and the optical
photometric period is shown in Figure 7.  Values from the spectra
of the 6 observatories are shown separately because the spectral
resolution differs among the instruments.  

Inspection of the H-\gal\ equivalent widths shows that they do not
correlate well with the optical photometric period.  While brief
sequences of agreement might occur [for instance  SAAO (Day 54-56), 
Magellan (Day 57--59), or MSSO (Day 61-65)], none of the 6
spectral sets of H-\gal\ measures is totally consistent 
with the photometric period determined over the same time interval.
We sought a periodicity in the SAAO spectra which has the longest
time span, but no meaningful period exists in the data.

The underlying assumption in the interpretation of the H-$\alpha$ profiles
holds that the emission arises from the accreting material as it
is channeled along magnetic field lines  from the disk to the star (Muzerolle et 
al. 2001; Natta \etal\ 2004), and fitting the profile with accretion
models gives an estimate of the mass accretion rate.  Thus we might
expect the equivalent width of H-\gal\ to exhibit behavior similar
to that of the X-ray accretion lines.  Inspection of Figure 7, in particular, during
the second and third \chandra\ pointings beginning Day 57 does not
show a convincing correlation. The flux of X-ray 
accretion lines decreases from Day 57.5 to Day 58.5, and the spectra from
SAAO and Magellan do show a decrease in the H-\gal\ flux. However,
beginning at Day 60.2, the accretion lines are at a high level
again (comparable to Day 57.5). While the Vainu Bappu
 spectra indicate a decrease in the equivalent width  of H-\gal\ between Day 60.2 and
 Day 61.3 similar to the change in the X-ray accretion line flux,  the 
Pico dos Dias spectra show a decreasing H-\gal\ flux between Day 60.7
and Day 61.6  when the 
X-ray accretion fluxes have not changed.   {\it We find no
correlation during the simultaneous observations (Day 57.5 -- 61.9)
among the H-$\alpha$ equivalent widths, 
the flux of X-ray accretion lines,  and the optical photometric
period.} Study of the line profiles themselves in the following
section are more  revealing.

\subsection{H-\gal\ and H-$\beta$ Profile Changes }
For 3 continuous nights (Day 57.51 to 59.92), MIKE spectra from 
Magellan/Clay with  
echelle resolution were obtained with 
high frequency (about 9 exposures of 100 seconds or less per
hour in addition to hourly longer exposures).  An additional
spectrum was taken on the 4th night.

Line profiles by their asymmetries can signal the presence of
differential mass motions  in the line-forming regions of 
optically thick lines (Hummer \& Rybicki 1968). 
In a static atmosphere, a line absorption coefficient remains at rest
centered on the profile. However with differential expansion, the
absorption coefficient moves to shorter wavelengths, weakening the
short wavelength side of the profile, and creating a `blue' side
that is weaker than the `red' side -- a red asymmetry.  And the opposite occurs when
differential inflow is present.  Thus the simple profile shape of 
these lines reveals the atmospheric dynamics in the line-forming region.
We seek such signatures of asymmetry in the line profiles.
 
The first night of the MIKE observations (Day 57.51-57.92) 
exhibited an unusual H-\gal\ profile (Figure 8, {\it left panel}) in that it has a
blue asymmetry (negative velocity peak is stronger
than the positive velocity peak) indicating inflowing
material. The H-\gb\ line (Figure 8, {\it right panel}), which is less optically thick
than H-\gal\, is striking in the inflow
signature of  blue asymmetry and a substantial absorption feature near $+$35~\kms. 
H-\gb\ occurs on both the  red and blue sides of MIKE and we have selected the
profile from the blue arm of the MIKE spectrograph since the line
is positioned near the center of the echelle order providing good 
photon statistics. Exposures for the blue side 
are generally longer (100 to 360 s) than for the red side 
which must be shorter so as not to saturate the H-\gal\ emission line. 
The H-\gb\ profile clearly shows a signature 
of inflowing material with its broad absorption
near $+$35 \kms\ that may correspond also to a feature in the H-\gal\ profile\footnote{The H-\gal\ region
has several water vapor lines which are discussed in Appendix A. They
do not affect our conclusions.}. 
Over the subsequent 3 nights of observation, the 
short wavelength wings of both H-\gal\ and H-\gb\ become systematically weaker.

The H-\gal\ profile during Night 1 (Day 57.51-57.92) of the MIKE
observations is unusual and not  
seen in the H-\gal\ profiles that we have assembled intermittently
over 7 years (2003-2010). This profile is broad, and relatively `flat-topped'
although the short wavelength peak is stronger than the long
wavelength peak (blue asymmetry).  Alencar \&\ Batalha (2002) in their
study of 42 H-$\alpha$ profiles spanning two years consistently 
report an asymmetric profile, much like Nights 2, 3, and 4 
reported here (Figure 8).  Straightforward interpretation of the
H-\gal\ profile suggests that the Night 1 profile indicates 
inflow ({\it i.e.} material moving towards the star) and
a stellar wind that is absent or extremely weak, only to become re-established during
the following nights.  The profile of H-$\beta$ confirms this
interpretation as it exhibits a strong inflow signature, and 
similarly increasing wind absorption each successive night.

The Night 1 H-\gal\ profiles in Figure 8  appear to represent 
the intrinsic stellar/accretion emission profile
unmodified by a stellar wind.  The red (long wavelength) and blue (short
wavelength) emission peaks are clearly correlated as shown in Figure 9.  This 
indicates that a single process produces  the  emission line and
suggests that the intrinsic H-\gal\  line profile is
approximately symmetric and broad. 
During subsequent nights, absorption from outflowing material sets in.
This systematic and uniform decrease of the short wavelength flux in
H-\gal\ is shown in Figure 10 where the ratio of normalized fluxes from a 1\AA\
bandpass on the short wavelength side (blue) to the long wavelength
side (red) is given for 3 nights.  During this time, the H-\gb\ line
behaves similarly. The positive velocity
component in H-\gb\ remains constant while the negative velocity side of the
H-\gb\ line becomes systematically weaker as the wind appears also in the 
less optically thick line. The abrupt change of the H-\gal\ blue:red
ratio during Night 1 is associated with an accretion event 
(Section 6 following)  and the profile becomes modified 
as the wind from the star develops absorbing the hydrogen
emission.  

The absorption on the short wavelength side of the H-\gal\ line yields an
estimate of the optical depth in the neutral component
of the stellar wind. 
At a velocity of $-$100~\kms\ in the H-\gal\ profile, the wind 
increases to  an optical depth of $\sim$0.5 by the  third night, assuming no
wind absorption on the first night. A lower
limit to the column density of hydrogen  required to produce this absorption
can be found with the assumption that T = 10$^4$K  in a wind that is
predominantly neutral hydrogen, and by taking the line absorption
coefficient to be thermally broadened.  These assumptions 
yield a lower limit of, $N_H \times L > 3 \times 10^{19}$cm$^{-2}$
in the wind from the polar regions of TW Hya.  This column density  is enhanced
in relation to a quiet sun coronal value of $\sim$2$\times 10^{19}$ cm$^{-2}$.

The source of the extremely broad H-$\alpha$ and H-$\beta$ profiles
has been  traditionally associated with the pre-shock accreting material.
Early models (Muzerolle \etal\ 1998a)  attributed a red-shifted H-$\alpha$
profile to its origin in the magnetically-confined accretion 
stream resulting in narrow H-\gal\ emission.  Later models (Muzerolle
\etal\ 2001; Kurosawa \etal\ 2011) incorporated line broadening 
and multidimensional non-LTE radiative transfer in the accretion 
stream, but the profiles remain pointed and
not similar to the profile of TW Hya during Night 1  
which is distinctly broader and flatter
than typical profiles for  classical T Tauri stars spanning a wide
variation in accretion activity (cf. Muzerolle \etal\ 2001, Figure 11).
Since the polar regions of TW Hya are facing us, redshifted emission
with velocities up to $\sim +$500 \kms\ (the expected ballistic
free-fall velocity assumed for the pre-shock material) would be
expected if the emission was produced by the infalling accretion
stream.  But this H-\gal\ profile is broad indicating plasma with
velocities both approaching and receding  up to  $\pm$150
\kms. Whereas  a pre-shock accretion stream
might account for red-shifted emission, it can not explain the blue-shifted
emission considering that TW Hya faces us pole-on. 
The full width at half maximum (FWHM) of the flat-topped H-\gal\ profile
during Night 1
ranges from 250 to 300 \kms; the thermal width of hydrogen 
at 10$^4$K is only 21 \kms.  Clearly the large values of the
line width are characteristic of 
supersonic turbulence in the region where H-\gal\ is formed.   
The profile requires the presence of a substantial turbulent 
component of 250-300~\kms\ to produce such broadening.  Note 
that the width of the turbulent component of the  \ion{Ne}{9} lines 
formed in  the accretion shock is also high, 
at 165$\pm$18 \kms\ (Brickhouse \etal\ 2010).  {\it Correlated strengths of
the blue and red side of the H-\gal\ line and the enhanced H-\gal\  
widths suggest that this emission (as well as H-\gb) is associated with
the subsequent cooling from the accretion shock and not from the
accretion flow} channeled from the circumstellar disk to the star.  The behavior
of these spectroscopic diagnostics with time discussed in the
following section can inform this model as well.

Studies of accreting T Tauri stars suggest that the width of the line at 
the 10\% level is  related to the accretion rate (White \&
Basri 2003; Natta \etal\ 2004; Curran \etal\ 2011).  Here, the width
of the H-$\alpha$   line at the 10\% level above a
continuum-subtracted normalized profile varies between 390 and 465
\kms, values that translate into an  accretion rate from 9.0$\times
10^{-10}$ to 4.2$\times 10^{-9}$ $M_\odot yr^{-1}$
according to the relationship proposed
by Natta \etal\ (2004). This interpretation relies on the assumption that
the entire H-\gal\ line arises from accreting material, and it 
is obvious from our observations that the wind plays 
a significant role in modifying these profiles.  The largest variation occurs on 
the short wavelength side of the profile, at negative velocities, and
is not consistent with motions expected from an  accretion stream.
Thus the values of accretion inferred from the widths may not be
meaningful.

\section{Accretion Events}

During the \chandra\ pointing, three `accretion events', signaled by an
abrupt short-lived increase in the flux of  the X-ray 
accretion lines occurred at Day 48.2, Day 57.54 (in 1 ks binning), and Day 57.74.
These are marked by arrows in Figure 5.  Details of these events and
the spectroscopic response are shown in Figure 11 (with the X-rays binned at 3 ks) and
Figure 12 (X-rays binned at 1 ks).

WASP photometry did not coincide with the first accretion event
and only one contemporaneous spectrum is available. 
The second accretion event is characterized by a sharp short (1 ks)
rise in the accretion-line flux by a factor of 2.
The third accretion event occurred $\sim$ Day 57.74 as marked
by an abrupt increase  ($\sim$26\%) in the flux of the X-ray accretion
lines that lasts for 3 ks. This event marks the 
highest hourly flux of accretion lines that occurred during the 
extent of the long 500ks \chandra\ observation.  

We first examine the 
long event on Day 57. Fortuitously, 
it occurred during intensive spectral
monitoring. As shown in Figure 11, the total H-\gal\ flux did
not change, but enhancement occurs  in the blue and red
wings of  H-\gal\ as well as a blue asymmetry in the line profile.  
Additionally the H-\gb\ flux increased, with the blue side of the 
H-\gb\ line showing greater enhancement (Figure 11, {\it panels D and E}).  

A detailed examination of the accretion  profiles reveals the time
sequence of this event. Inspection of the 1ks binning of accretion 
lines (Figure 12) shows the abrupt increase starts
at 05:46 UT (Day 57.74), where we take the midpoint of the 1ks bin.  The H-\gal\
asymmetry changes after 05:55 UT, 9 minutes later.
The H-$\beta$ line also exhibits an increasing blue asymmetry peaking
slightly after the H-$\alpha$ line. {\it The sequential and abrupt profile
changes following X-ray accretion suggests H-\gal\ and H-\gb\
originate in the downflowing post-shock cooling region.} These changes 
provide strong evidence too that the
increase of the X-ray accretion component represents an 
actual increased accretion
rate and not a decrease in the column density of accreting
material. The latter process could  reduce the soft X-ray absorption and 
thereby strengthen the observed flux of the accretion lines.
The veiling (discussed in Section 9) also increases following this
event, but delayed by $\sim$ 2 hours. 

H-\gal\ profiles 
taken before (Day 57.66) during (Day 57.79) and at the end
of the event (Day 57.92) are shown in Figure 13.  
All parts of the profile behave
similarly, suggesting that  the emission measure of a turbulent 
region increased.  And considerable variability occurs 
in the total H-\gal\ flux as well.   

The helium line profile (Figure 14) changes quite
dramatically beginning at the same time as the H-\gal\ asymmetry
change.  The helium line is known to have a broad and narrow 
component and the broad component is frequently identified as
an accretion signature (Muzerolle \etal\ 1998b; Alencar \& Batalha
2002; Donati \etal\ 2011). Before a significant increase in the total line flux,
the profile begins to broaden, particularly on the positive velocity
wing  (between +100 and +200 \kms).
The \ion{He}{1} (\gla\ 5876) line has not changed in flux at 06:16 UT
(Day 57.762), however, the narrow line component increases 
and the wing broadening appears in the
spectrum at 06:18 (Day 57.763, about 30 minutes after the increase in the
X-ray accretion lines).  The flux of the 
\ion{He}{1} 5876\AA\ line (Figure 12) increases 
about 2.5 hours after the X-ray accretion event.
Decomposing the profiles into a narrow and broad component, we find, at the end of the
observing night as compared to the profile earlier in the night, 
that the narrow component remains at constant flux  (to within 15 percent) 
and the FWHM to within 1 \kms, while the broad component 
increased both in flux by 30\% and width by  7 \kms, and its  redshift
increased from $+$15 to $+$21 \kms\ with respect to the narrow component.  
Interpretation of this redshift must be treated with caution because
increased short wavelength absorption can result from outflowing
material ({\it cf.} Section 8) and cause 
an apparent redshift in line position. 
The broad (gaussian) component has a FWHM of 175
\kms\ on average.  This width exceeds the thermal width
of helium at 10$^4$K (11 \kms), and is similar to  the wide
lines observed in \ion{Ne}{10} and H-\gal.  {\it Therefore we also associate
the broad component of \ion{He}{1} with the turbulent post-shock
region.}

The earlier short accretion event (Day 57.54) shows X-ray enhancement
for a  1 ks interval.  The H-\gb\ flux and the red and blue segments of H-\gal\
are decreasing at the start of the optical observations, but
we have no knowledge of prior activity in the X-ray emission. The
ASAS photometry (Figure 4) hints that flaring may have occurred
near Day 57.66. Subsequently, and quickly,
the veiling increases sharply, followed by an increase in
the flux of the helium line.  This event exhibits a substantially
smaller impact on the spectroscopic diagnostics than the later
accretion event discussed above, perhaps because of its shorter
extent in time.

\section{Coronal Flare Event}

A  flare occurred at Day 58.2 (Figure 15).
The H-$\alpha$\ profile was measured once during the flare at the SAAO,
and the line flux was comparable to earlier values.  
Spectra over the next 5 hours exhibited a decrease in the H-\gal\ 
equivalent width by $\sim$8\%, attributed to the decrease in strength of the 
short wavelength side of the profile, similar
to that shown in Figure 8 between Day 58.5 and 58.92. 
The WASP optical photometry suggests flaring events occurred
during  Day 55, and the SAAO photometry indicates that TW Hya
was brighter then too. However the SAAO H-\gal\ equivalent width
remains low during that event.

We conclude that  {\it the  flare affected the V-band photometry 
of TW Hya causing an optical brightening (a 'white-light flare') 
but does not change the flux nor the characteristics
of the H-\gal\ profiles.  This flare appears to be a stellar
coronal event unrelated to the accretion process.}  

\section{The Wind in Helium $\lambda$10830}

The near-IR \ion{He}{1} 10830\AA\ line is formed in the stellar
atmosphere and wind, and its metastable nature allows a good probe
of stellar winds  especially in T Tauri stars  (Edwards \etal\ 2003;
Dupree \etal\ 2005).   One observation in 2002 (Dupree \etal\ 2005)   
showed the wind extending to a speed of $-$280~\kms;in 2005 the
terminal velocity reached $\sim$300~\kms. A spectrum obtained on 1
March 2007 (Figure 16) with PHOENIX at Gemini-S  shows 
that wind opacity is larger than previously observed, particularly at
velocities greater than $-$200~\kms\ and the terminal velocity is slightly higher
($\sim$325 \kms).  The emission component of \gla10830 is substantially weaker
as well.  In addition, 
subcontinuum absorption at positive velocities is stronger then
measured previously and
extends to $+$300~\kms.  An optical spectrum taken at Magellan 60 minutes
prior
to the PHOENIX observation is instructive (Figure 18).
Not only are the H-\gal\ and H-\gb\ lines
weakened from the previous 3 nights (cf. also Figure 8), but 
the profiles are characterized by wind absorption.
In fact, the \ion{He}{1} \gla5876 transition and H-\gb\ both
exhibit a narrow wind absorption feature at $\sim -$125 \kms.
The \gla5876 transition arises from the upper $^3$P level
of the \gla10830 near-IR line, and absorption by the wind
is similar to that found in the \gla10830 profile, although
not extending to such high velocities. 
Higher speeds observed in the near-IR \ion{He}{1} line than H-\gb\ 
suggest an accelerating outflow.    
Clearly, both the wind speed and opacity vary with time, but insufficient
observations prevent a direct association of the near-IR \ion{He}{1} 
line profile with the accretion process.

\section{Veiling}
 
Veiling, which describes the addition of
a continuum spectrum to the stellar spectrum, is thought
to occur in TW Hya due to the presence of an accretion
`hot spot' (Alencar \& Batalha 2002).  Magellan spectra taken during
the nights of 26 and 27 February 2007 (Day 57 and 58) were used to
determine the veiling because they were simultaneous with
the \chandra\ pointing.  The T Tauri star 
spectrum is modeled as a normal stellar photosphere plus a smoothly
varying veiling continuum.  The normal photosphere is taken
from a  spectrum of GJ1172, a K7V star obtained at Magellan (1200 s exposure)
during the February campaign.  This template
star was used previously by Alencar \& Batalha (2002) who
state that it ``presented a perfect match to the photospheric
lines of TW Hya, after it was rotationally broadened to the TW Hya
$v sin i$ and artificially  veiled.'' The MIKE spectra of GJ1172 when 
rotationally broadened with  $v sin i = 5$ \kms, confirm that
it is a good match for TW Hya.

We follow the procedures outlined by Hartigan \etal\ (1989)
by constructing a rotationally broadened normalized  spectrum
of GJ 1172 to use as a template.  Many spectral regions, 
ranging in width from 5 to 10\AA\
were selected from the centers of echelle orders on the 
blue and red side of the MIKE spectra.  The observed 
normalized TW Hya spectrum,
O($\lambda$)/O($\lambda$)$_{cont}$,
is related to the normalized template spectrum,
T($\lambda$)/T($\lambda$)$_{cont}$:

\begin{equation}
\frac{O(\lambda)}{O(\lambda)_{cont}}=\frac{[\frac{T(\lambda)}{T(\lambda)_{cont}}+r]}{[1+r]}.
\label{eq1}
\end{equation}

\noindent
where $r$ represents the fractional veiling 
that is  produced by the veiling
continuum, $V$: 

\begin{equation} 
r=\frac{V(\lambda)_{cont}}{T(\lambda)_{cont}},
\label{eq2}
\end{equation}

By changing the value of
$r$, the right side of Eq. 1 becomes an 
artificially veiled template spectrum to be compared to 
the observed TW Hya spectrum  yielding the veiling in the 
selected regions.   Iterations  through $r$ values in 
steps of 0.1, allow the best fit to be determined by seeking
a minimum value of $\chi^2$ between the template and the TW Hya
spectrum.
Values of the veiling ($r$) as a function of
wavelength are shown in Figure 18 for 5 spectra spanning 5 hours.  The
increased veiling towards shorter wavelengths is well known (Alencar
\&\ Batalha 2002) and expected from a continuum source with T$\approx$8000K.

During this time, the veiling varies, which appears to result
principally from the changing size of the accretion hotspot.
In order to search for a relation between the veiling values and the
X-ray flux, we introduce an average value of the veiling based on seven
measurements between 4400\AA\ and 5000\AA\ taken from the same regions
in all spectra.   These are the values used
previously in Figure 12; the complete set is given in Table 3 and   shown in
Figure 19 ({\it upper panels}) aligned with the simultaneous measures of
the total X-ray flux.  The veiling values and X-ray fluxes were
iterated through positive time shifts from 0 to 0.5 day in increments of 0.01 day. The 
correlation between the veiling and the
total flux is strongest when the veiling measures are shifted by $+$0.11~d 
({\it lower panels}) as shown in Figure 19, giving a correlation
coefficient of 0.78.  The correlation between 
the veiling parameter and the simultaneously measured total X-ray flux is
shown in Figure 20.

The principal source of the total X-ray flux is most likely the corona. 
Our models based on the  \chandra\ spectrum (Brickhouse \etal\ 2010) indicate
that most of the total X-ray flux arises from the corona.  In order to
confirm this conclusion, we selected only the 
highest energy component of the
total flux, namely that greater than 1 keV, to correlate with the
veiling.  The same delayed shift of 0.11~d gives the highest
correlation coefficient, with a values of 0.72, only slightly lower
than the correlation with the total flux. 
This suggests that {\it the
corona responds to the  increase in  veiling produced $\sim$2.5 hours
earlier.} We searched for similar correlations between line components of the
X-ray spectrum itself without definitive results.    Taking the line spectrum
divided into accretion and coronal components  does
not yield a significant correlation with appropriate lag between
the components themselves and the veiling.  For this, we evaluated small negative as well as
positive lags between the accretion line flux and the veiling values.
Because the counts are low in the accretion and coronal line
components and affected by the underlying continuum, any correlation
would be weakened. 

\section{A Suggested Model}

These results suggest that the accretion process can be traced 
spectroscopically in the 
following way.  Material channeled along the magnetic flux tubes threaded from the
circumstellar disk accelerates to supersonic velocities and creates a 
shock near the stellar surface (Figure 21).  Brickhouse et al. (2010) identified
emission features in the \chandra\ X-ray spectrum that originate
in the MK shocked plasma. The shock size  as   estimated
from the emission measure and electron densities derived from the
X-ray spectrum has a scale length 
of 3.2$\times$10$^4$ km. The shocked material cools in a turbulent
plasma as it approaches the stellar photosphere and the profiles 
of the optical emission lines mark the progress of the event.
Following the major accretion enhancement (Day 57.74) noted in Figures 11 and
12, the H-\gal\ and H-\gb\ line profiles show a signature of abruptly 
increasing downflow (towards the star) revealed by the increase in 
blue:red asymmetry. It is important to note that both Balmer series
lines are broad, and the line wings are roughly symmetric.  The FWHM of
these lines reaches $\sim \pm$150 \kms\ which indicates substantial 
turbulent broadening in excess of  the thermal width ($\sim$21~\kms), and 
is comparable to the turbulent velocity inferred from the breadth of the
\ion{Ne}{9} lines in the post-shock accretion zone, namely 165 \kms\   
measured in the \chandra\ spectrum (Brickhouse et
al. 2010). Additionally, hydrogen emission lines in the infrared 
(Vacca and Sandell 2011) display broadened line wings amounting to
several hundred km s$^{-1}$, consistent with the H-\gal\ profiles.

The \ion{He}{1} \gla5876
transition exhibits an increase in the broad wing component after 
the downflow signatures occur in the Balmer lines.  This broadening 
occurs over a time scale of minutes (Figure 14).  The veiling and the \ion{He}{1}
flux increase about 0.08 day ($\sim$2 hours) after the accretion event
marked by X-rays.  Such a time scale is roughly consistent with
the time scale suggested by the inflow velocity and size of the hot
spot.  The size of the  photospheric `hot spot' on TW Hya has been estimated to
cover 2-6\% of the apparent stellar disk, although at times
it extends from the pole to lower latitudes and the area
coverage increases to $\sim$25\% (Costa \etal\ 2000; Batalha \etal\
2002; Donati \etal\ 2011). Simulations of disk accretion 
(Romanova \etal\ 2004) suggest coverage values ranging from 1 to 20\%.
We take 10\% for this example, and for a stellar radius of
0.8R$_\odot$, the hot spot has a radius of 3.5$\times$10$^5$ km.
For a velocity of 35 \kms\ corresponding to the H-\gb\ 
absorption feature, this  suggests a characteristic time scale of 2.8
hours which is in harmony with the observed change in the veiling.
Following the changes in photospheric veiling, the coronal X-rays
respond to this increase with a delay on average of 0.11 days
(2.6 hrs, Figure 20).  These results give strong evidence for an accretion-fed
corona which may contribute to the wind acceleration as well.

It is tempting to attribute the absence of a wind followed by the
increase in wind opacity as a consequence of the accretion event. One
could envision a dramatic change in the magnetic field configuration 
caused by a sharp increase in the amount of accreting material,
allowing a wind to develop.  However, the 
rotation of the star over several days could also present a different
aspect of an asymmetric accreting geometry and outflow.
  
\section{Comparison with Previous Results}

Our model for TW Hya suggests the contribution to H-\gal\ from the
post-shock cooling region must overwhelm any contribution from 
the accretion stream. Previous models have not considered the 
contribution from the chromospheric or post-shock material.  
Muzerolle et al. (2000) assume that all of the H-\gal\ emission 
arises in the  hot stream of material extending  
out to a few R$_\star$,  channeled from the circumstellar disk 
by the magnetic field, and free-falling to produce the accretion 
shock.  Another paper (Muzerolle et al. 2001) in more detail, 
explicitly shows  the geometry that is generally assumed.  
These papers do not include a contribution to the H-\gal\ profile 
from the post-shock cooling zone or the chromosphere.  
A  simple calculation demonstrates that the post shock/chromospheric 
region has substantially more material than the plasma in the 
accretion stream, and most likely will dominate formation of 
the H-\gal\ profile.

Mass conservation (Brickhouse et al. 2010) yields the preshock 
electron number density (5.7 $\times$ 10$^{11}$  cm$^{-3}$), and the 
filling factor of the shock (1.1\% of the stellar surface) based 
on a dipole magnetic field (Cranmer 2008). Assuming a truncation 
radius for the gas disk of 4.5 R$_\star$ where R$_\star$ = 0.8
R$_\sun$, we  estimate an upper limit to the emission measure 
of a (cylindrical) channeling tube  of length 3.5 R$_\star$ as 
$N_e^2 \times V = 2.8 \times 10^{55}$ cm$^{-3}$ where $V$ is the
emitting volume.  This 
is  an upper limit for several reasons.  The density falls 
off towards the disk by a factor of 10 in the Muzerolle et al. (2001)
models,  and the accretion streams have a length of  only 1.2 to 2 R$_\star$,  
decreasing the emission measure  to $\sim 2 \times 10^{54}$ cm$^{-3}$ or less.

For comparison, what is the expected  chromospheric emission measure
resulting from the accretion shock?  Modeling  gives the post-shock
electron density as $2.5 \times 10^{15}$ cm$^{-3}$ at the bottom 
of the chromosphere where the accretion stream ram pressure 
matches the gas pressure. We take the  density of  
$7.7 \times10^{13}$ cm$^{-3}$, which is the mean between the
immediate post shock density and the value at the base of the
chromosphere,   and a length scale of half 
the post-shock cooling distance, namely  200
$km$, (Brickhouse et al. 2010).  The area of the
chromospheric/photospheric hot spot appears to be  $\sim$3\% (Donati et
al 2011) of the stellar surface (1.2 $\times$ 10$^{21}$  cm$^2$ for a star with
R=0.8R$_\sun$).  The emission measure $N_e^2 \times V \sim  1.4 \times
10^{56}$ cm$^{-3}$ is a factor of $\sim$100  larger than 
the emission measure from the accretion stream. Another  
estimate of the emission measure of the hot region at the accretion
footpoints responsible for the near-infrared 
hydrogen Paschen series emission in TW Hya  
(Vacca and Sandell 2011) gives an emission measure 
of $0.7-4 \times 10^{55}$ cm$^{-3}$, a value larger than from the
accretion stream and compatible  with the
characteristics derived above for the post-shock cooling region.  

What might be the effects of a deviation from a dipole magnetic
configuration? Donati \etal\ (2011) suggest that the star may have an
octupolar component  to the magnetic field in the photosphere, but an
octupole field falls off exceedingly rapidly (r$^{-5}$) with distance from
the star as compared to a dipole (r$^{-3}$) and is generally thought not
to extend  far enough to penetrate the circumstellar disk.  In any
case, the octupole field suggested by Donati \etal\ has a pole (roughly)
coincident with the stellar pole where accretion is believed to occur.
Thus if accretion towards the pole proceeded on a octupole field line,
it would be hard to differentiate from a dipole configuration.
However in our view, the accretion stream is not the source of the
H-\gal\ profile, as discussed previously.

The observed turbulent motions are compatible with cooling zone
parameters derived from the X-ray observations. Densities in the 
cooling zone range from 2$\times$10$^{15}$ to
2.3$\times$10$^{12}$cm$^{-3}$ and temperatures from several MK 
in the immediate post-shock region. down to $\sim$10$^5$ K. 
Donati et al (2011) suggest that the octupolar component 
of the photospheric field at  the pole reaches a maximum of  
2.5-2.8 kG, whereas the dipole field varies from 0.4-0.7 kG. 
In the photosphere, the magnetic pressure caused by the 
octupole field would  dominate the gas pressure at 
an effective temperature of 4000K.  An octupole field 
falls off rapidly (r$^{-5}$) and this octupole field 
strength becomes commensurate with the dipole 
field at 2R$_\star$. For heights greater than 0.2R$_\star$ above 
the photosphere, at a temperature of 10000K  in a 
dense post-shock region, the gas pressure becomes comparable 
to the magnetic pressure from an octupole field.  Other 
measures of the magnetic field (Yang et al. 2007) 
suggest the field is substantially smaller than indicated 
by the Donati et al (2011) modeling, and  pressure balance 
would occur even closer to the photosphere.  However, if MHD 
turbulence is present, the strong fields do not impede 
large motions in the plasma, and  the observed turbulence 
could be present at any level above the photosphere.  

We note that accretion shock modeling to date has been unable 
to reproduce both the profiles and the intensities of the 
observed ultraviolet lines from TW Hya (G\"unther and Schmitt 2008; 
Lamzin et al. 2007). The emission measures of the post-shock regions 
discussed here are large, much larger than found in a normal K7 star, 
and would be expected to form a strong H-\gal\ line.    We can
estimate the expected H-\gal\ flux  based on the observed Paschen 
series lines which terminate on the n=3 level of hydrogen, the upper 
level of the Balmer H-\gal\ line. A sequence of lines has been 
measured in TW Hya (Vacca and Sandell  2011), but this does 
not include the strongest line in the series, the Paschen-\gal\ 
transition.  Assuming the total number of photons 
landing in the n=3 level  produces the H-\gal\ line, we find 
an H-\gal\ flux  of  7.7 $\times$ 10$^{-12}$ erg cm$^{-2}$s$^{-1}$ 
which represents a lower limit because the contribution of the  
Paschen-\gal\ line is not included.  Rucinski and Krautter (1983) 
estimate the H-\gal\ flux from TW Hya to be: 1.6 $\times$ 10$^{-11}$erg cm$^{-2}$s$^{-1}$ 
which is in harmony with the value from the infrared emission.   
Thus the post-shock material inferred from the Paschen series 
(Vacca and Sandell 2011) can produce the observed H-\gal\ flux. 
The size of this region depends on the density.  For a temperature
$\ga$7500K,  
a density of 10$^{12}$ cm$^{-3}$, and a 4\% covering 
fraction, Vacca and Sandell (2011) find that the  thickness 
ranges between  10$^4$ to 10$^5$~km; this value decreases 
to 340--820~km if the density is an order of magnitude larger. 
Although current theoretical models of post-shock conditions 
generally focus on temperatures producing X-ray emission, 
a 1-D post-shock  model (Sacco et al. 2010) contains 
chromospheric regions at temperatures of 8000K with a thickness 
of $\sim$1500km, and densities of 10$^{15}$ cm$^{-3}$. 
This is denser and may be  thinner than indicated by the 
Paschen series measures.  Additionally the models remain
static at these chromospheric temperatures in contrast to 
the motions inferred from the changing profile shapes observed in H-\gal.  

The  observations reported here suggest that the broad (symmetric) 
H-\gal\ line arises from a large turbulent region that 
appears to be  part of the post-shock cooling region. 
Such an H-\gal\ volume is not considered in current 
model calculations and these observations can 
empirically define its characteristics.
It may be associated with the large  \ion{O}{7} region 
that is extrinsic to the currently postulated post-shock region.

\section{Conclusions}

This comprehensive study clarifies the accretion process, and
demonstrates the power of simultaneous spectral diagnostics and 
photometry to probe the plasma dynamics in TW Hya.  We find: 

\begin{enumerate}   
\item{The periodicity in the optical photometry appears unrelated to the H-\gal\ flux, the coronal
X-ray flux, and the accretion line flux suggesting an origin distinct
from the accretion process.  While assignment of
these variations to short-lived hot spots on the stellar surface
has been offered  by many (Hu\'elamo \etal\, 2008, Mekkaden 1998; Batalha \etal\
2002), other authors (Rucinski
\etal\ 2008; Siwak \etal\ 2011) have conjectured that changing optical periods may be
associated with Keplerian rotation of 'blobs' of large magnetically 
controlled structures in the circumstellar
disk. Our observations appear to eliminate photometric variability directly
associated with the accretion process; we can not discriminate among
other possibilities.  }

\item{The H-\gal\ profile appears to be intrinsically symmetric and the
 short and long wavelength sides of the emission are correlated. We
suggest this broad line and H-\gb\ as well  arise in the turbulent post-shock cooling
region, and not, as frequently assumed from the funnelled accretion
flow from the circumstellar disk to the star as a `ballistic infall signature.' 
The H-$\beta$ line appears to be a superior diagnostic of
 dynamics because it is less optically thick than H-\gal.  Both of
these profiles are substantially modified by a wind originating from the
star that increases in optical depth over several days. These changes
may be driven by the X-ray accretion  or the rotation of the
star presenting different aspects of an asymmetric outflow over
the polar regions.  Modeling of the optical line profiles must
recognize the complex origin of the emissions, and  modifications
by the stellar outflows.  The presence of a turbulent
cooling region offers a straightforward explanation of the broad
 permitted
emission lines including those in the UV and far-UV regions
(Herczeg \etal\ 2002, Dupree \etal\ 2005).} 

\item{The temporal variation of the \ion{He}{1} 5876\AA\ line profile in addition
to its correspondence in width with other species suggests  the
post-shock  cooling region also produces the broad component
of the profile.}

\item{This study clearly reveals the sequence of the accretion process
in various energy bands and spectral lines. Enhanced accretion,
inferred from the X-ray accretion line fluxes is followed  
by inflow signatures in H-\gal, H-\gb,  and an increased flux in
the  broad wing of the \ion{He}{1} (5876\AA) line
arising in the turbulent post-shock cooling region. Subsequently the photospheric 
veiling increases, and the response of the stellar corona  
2.5 hours later, follows the increase in the photospheric veiling.}

\item{Spectroscopy gives direct evidence for the influence of the
 accretion process in heating the corona 
and provides observational confirmation of simulations of
coronal heating and wave-driven winds (Cranmer 2008, 2009) .}

\end{enumerate}

\acknowledgments
 We are grateful to Jonathan Irwin for assistance with the WASP data.
The WASP consortium comprises the University of Cambridge, Keele
University, University of Leicester, The Open University, The Queens
University Belfast, St. Andrews University, and the Isaac Newton
Group.  Funding for WASP comes from the consortium universities and
from the Science and Technology Facilities Council of the UK. We
appreciate
the efforts of Luca DiFabrizio who reduced the TNG/SARG spectrum. This
paper includes data gathered with the 6.5 meter Magellan Telescopes
located
at Las Campanas Observatory, Chile. 
Also, based on observations obtained at the Gemini Observatory, which is operated by the 
Association of Universities for Research in Astronomy, Inc., under a cooperative agreement 
with the NSF on behalf of the Gemini partnership: the National Science Foundation (United 
States), the Science and Technology Facilities Council (United Kingdom), the 
National Research Council (Canada), CONICYT (Chile), the Australian Research Council (Australia), 
Minist\'erio da Ci\^encia, Tecnologia e Inova\c{c}\~ao (Brazil) 
and Ministerio de Ciencia, Tecnolog\'ia e Innovaci\'on Productiva (Argentina).

{\it Facilities:} \facility{ASAS (U. Warsaw)},\facility{CXO (HETG)},
\facility{Gemini:South (PHOENIX)}, \facility{KECK (NIRSPEC)},
  \facility{Magellan:Clay (MIKE)},
\facility{SAAO:1.9m(Cassegrain spectrograph)},
\facility{SSO:2.3m(echelle)},  
\facility{Pico dos Dias:1.6m (coud\'e)}, \facility{TNG (SARG)}, 
\facility{Vainu Bappu:2.3m(echelle)}, \facility{WASP-South}.

\clearpage

\begin{figure}
\includegraphics[angle=0.,
scale=1.]{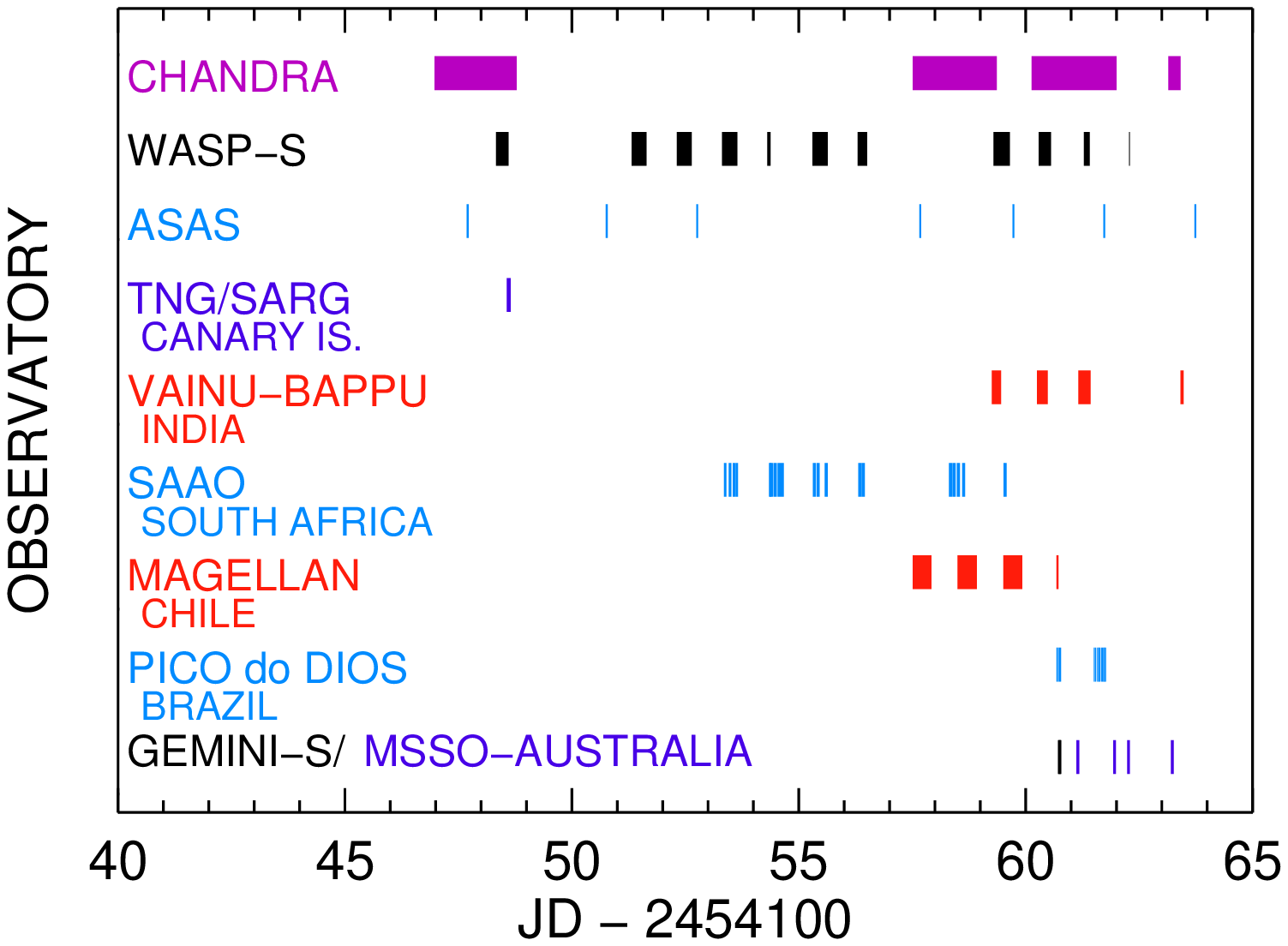}
\caption{Overview of the dates of observations at the participating
observatories and the \chandra\  pointings on TW Hya.}
\end{figure}
\clearpage

\begin{figure}

\includegraphics[angle=90.,scale=0.7]{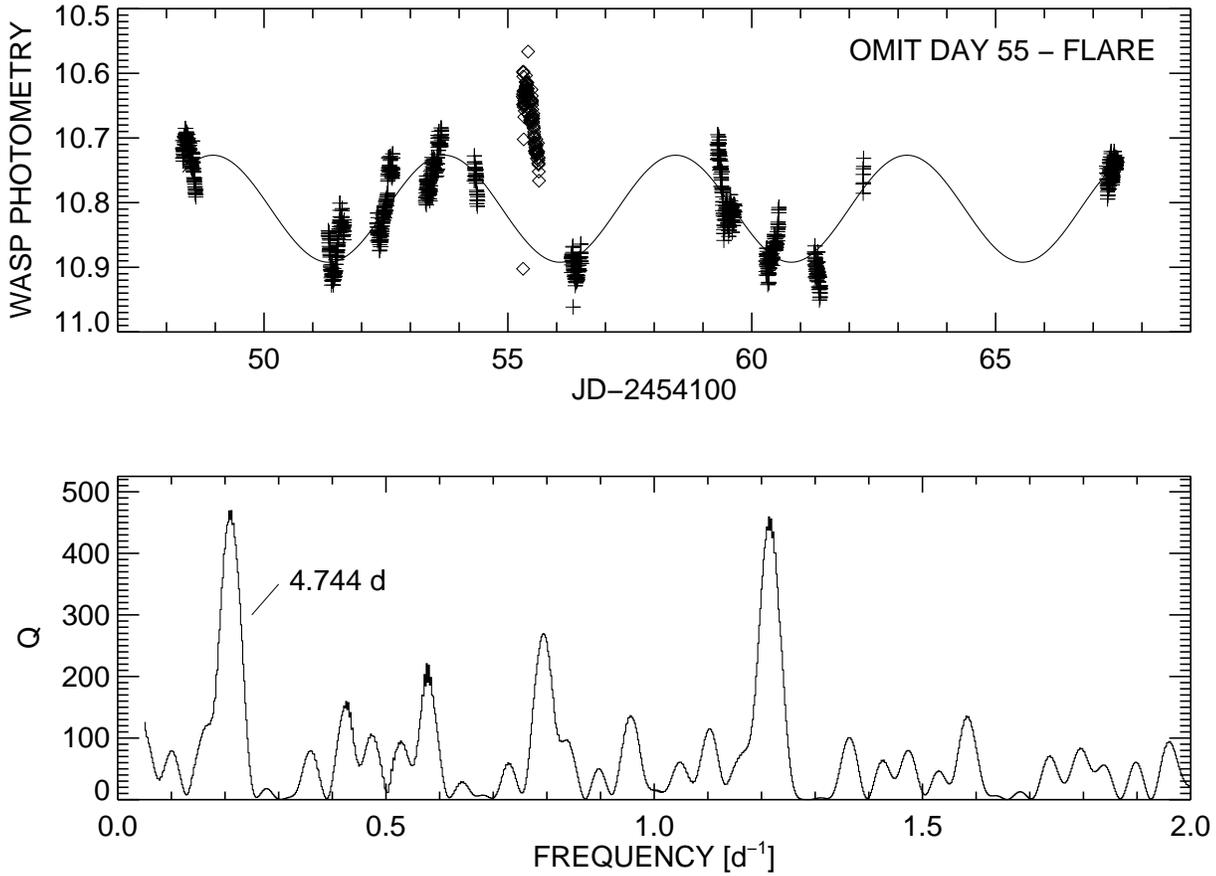}
\caption{{\it Upper panel:} Light curve of TW Hya from WASP-S measured
typically every 120~s.
Errors in the magnitude per bin range from 3 to 5 mmag.  A substantial
flare occurred during JD 2454100+55  (points denoted by open diamond symbols) 
and this night has been omitted from the
period search.  A sine curve
with the derived period (see below) overlays  the data. {\it Lower
  panel:} The periodogram reveals one strong period during this time at 
  a frequency of 0.210800 (d$^{-1}$) corresponding to 4.744 d.  The
peak at 1.21 d$^{-1}$ corresponds to an alias of +1 cycle d$^{-1}$.}

\end{figure}
\clearpage


\begin{figure}\vspace*{0.3 in}
\begin{center}
\includegraphics [angle=90.,
  scale=0.7]{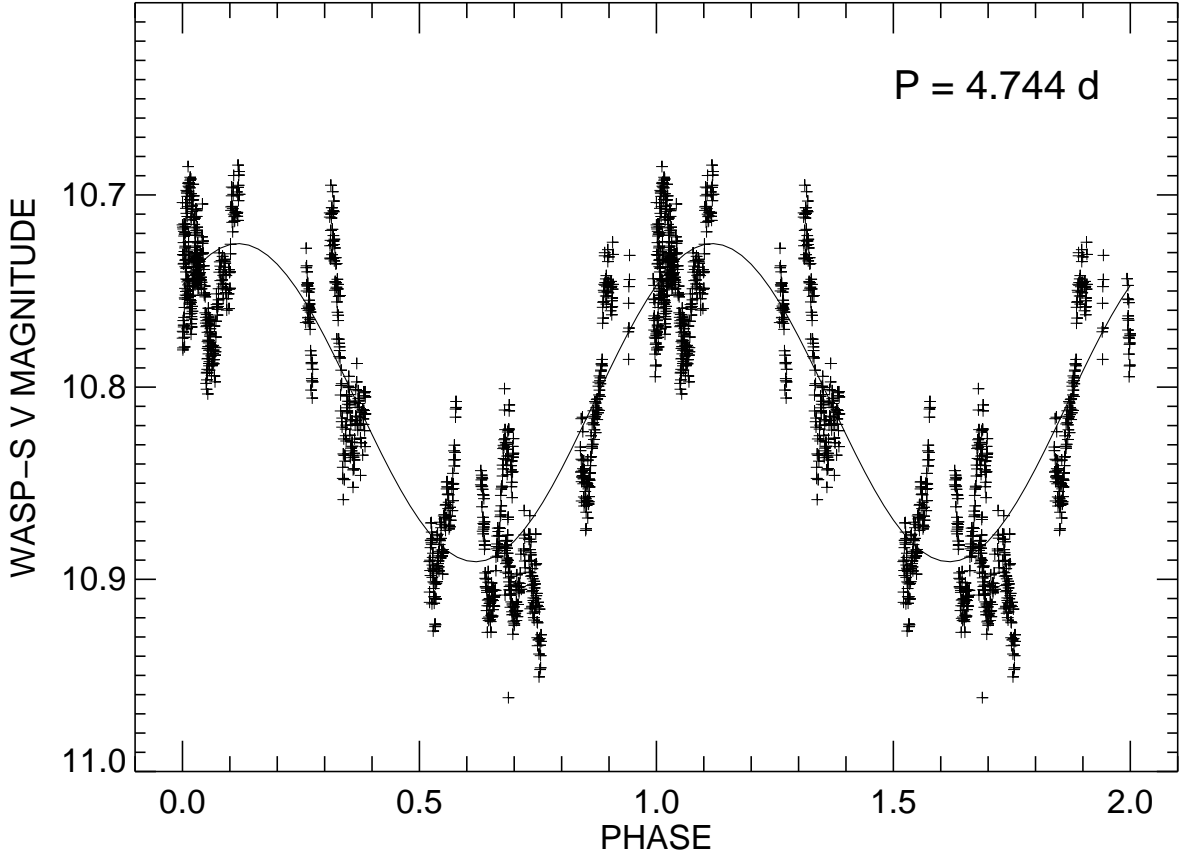}
\caption{The WASP-S photometry phased to the period of 4.744 d where
the flaring episode identified in Figure 2 has been omitted.  The data
are 
repeated twice in this figure. 
These magnitudes should be better than 1\% for a star as bright as
TW Hya. The substantial real  
short term variability is similar to that found in other T Tauri
stars.  }
\end{center}
\end{figure}
\clearpage

\begin{figure}
\begin{center}
\includegraphics[angle=0.,scale=0.8]{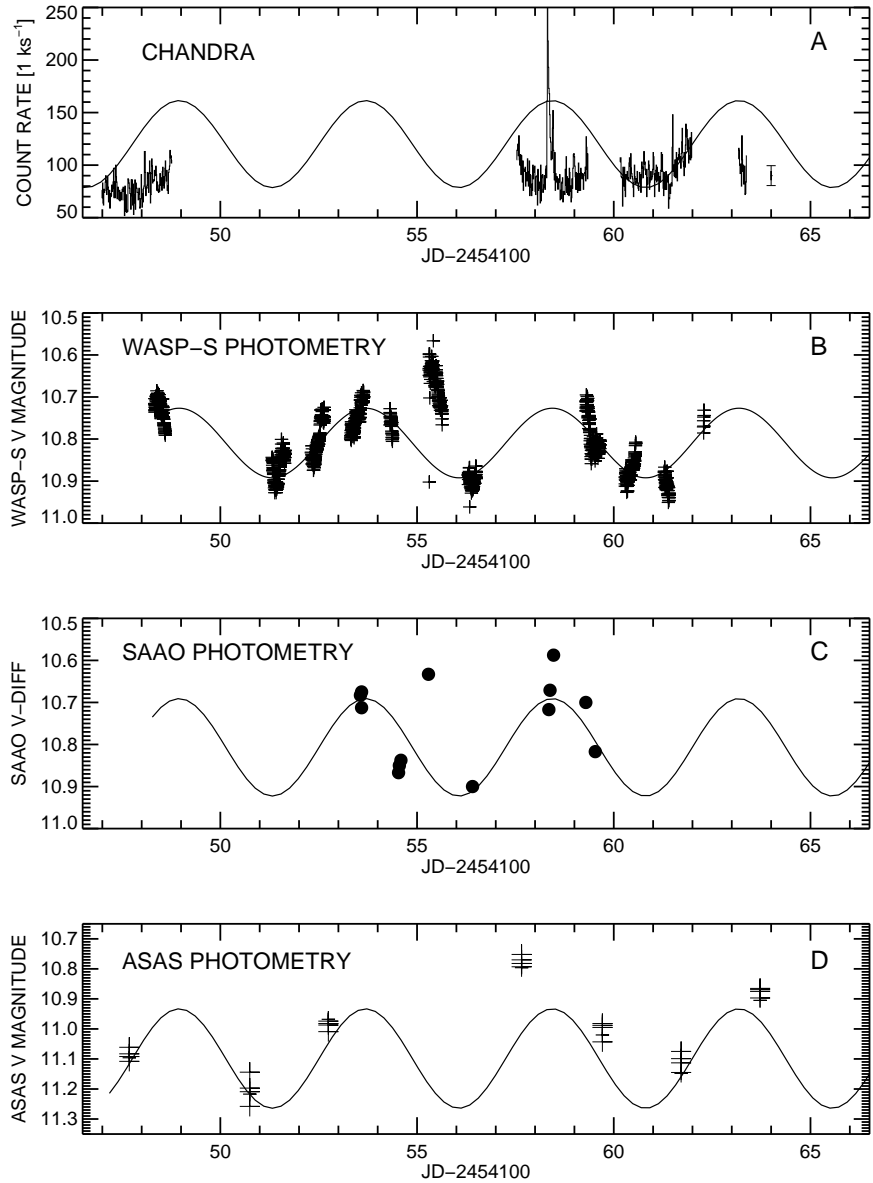}
\caption{CHANDRA first-order (coronal) X-ray light curve  binned to
  1ks ({\it Panel A})  
and contemporaneous photometry: WASP-S ({\it Panel B}), 
SAAO V-band differential photometry ({\it Panel C}), and ASAS
  photometry from five apertures, 
({\it Panel D}). A sine curve with the 4.74 day period
derived from WASP-S photometry with 
arbitrary amplitude overlays the \chandra\ flux and the SAAO and ASAS photometry.
Two white-light flaring episodes appear as indicated by the photometry during Day 55 and 
Day 57--58; the latter coincides with the \chandra\ (coronal) X-ray flare.}
\end{center}
\end{figure}

\clearpage


\begin{figure}
\begin{center}
\includegraphics[angle=0.,scale=0.8]{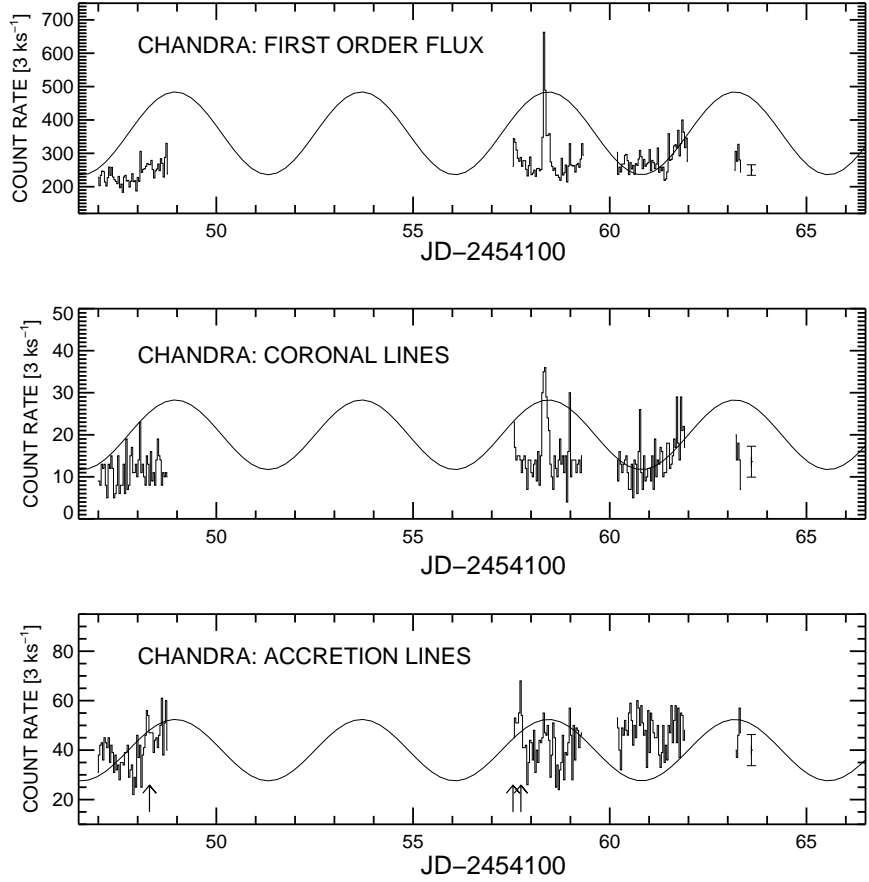}
\caption{CHANDRA first-order light curve  ({\it top panel}) containing
lines and continuum   
and a separation of the emission line X-ray spectrum into
features arising from the   hot stellar corona ({\it middle panel}) and 
the accretion shock  ({\it lower panel}). All data are binned over 3~ks. 
A sine curve with the 4.74 day period derived from contemporaneous
photometry,  and  
arbitrary amplitude  is overlaid in each panel to 
aid the eye. It is easy to see from the coronal lines ({\it middle
  panel}) that the X-ray flare ({\it top panel}) during Day 58 is
predominantly a coronal event as found by Brickhouse \etal\ (2010).  
The accretion lines vary in flux by a
factor of 1.5, and two localized 'accretion events' (marked by arrows
in the {\it lower 
panel}) occur at the same photometric phase.}
\end{center}
\end{figure}
\clearpage
\begin{figure}
\begin{center}
\includegraphics[angle=90.,scale=0.7]{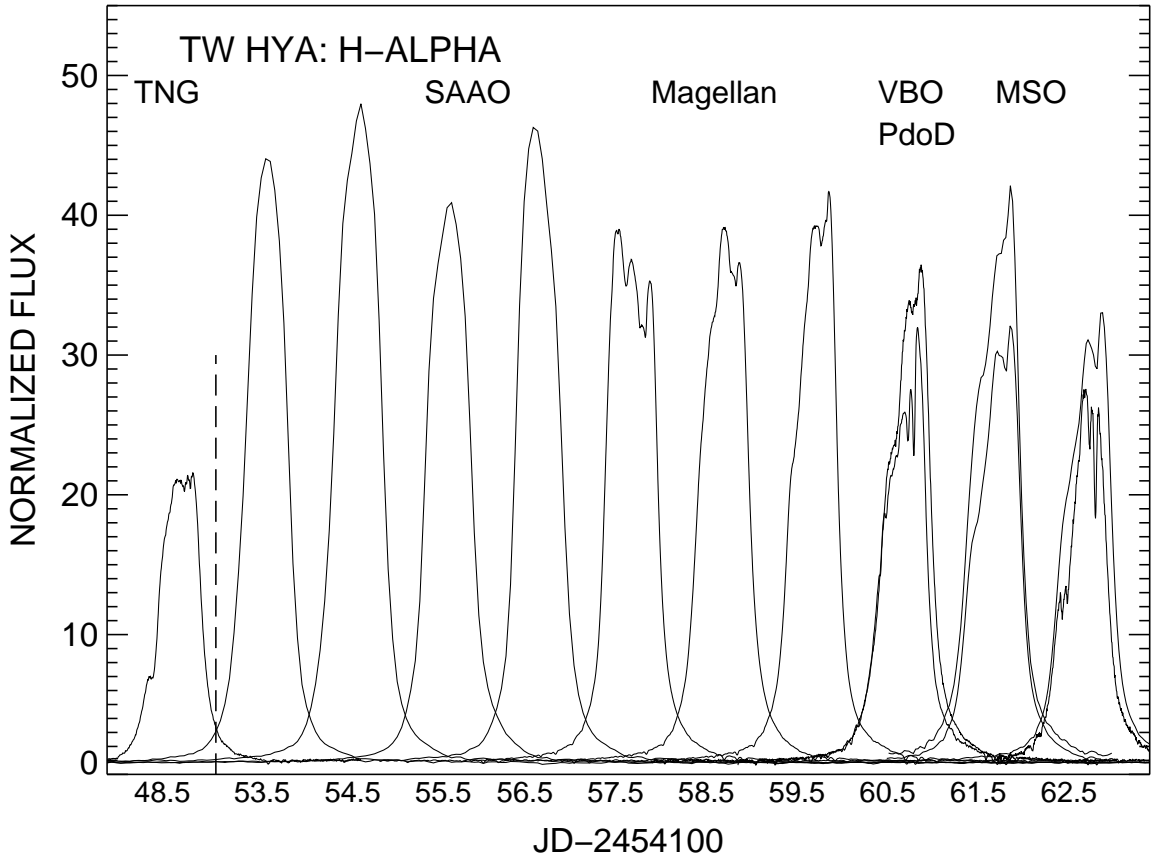}
\caption{Nightly averages of  H-\gal\ profiles. The spectroscopic
  resolution and the number of spectra per night both vary
  substantially. The lower resolution profiles from SAAO (Day 53.5-56.5) have
  been multiplied by 0.5 to normalize them to the Magellan spectra
taken at the same time on Day 58.5. Note the gap in the X-axis between the first and second profile
  (marked by a broken line).  The facilities used are marked.}
\end{center}
\end{figure}
\clearpage

\begin{figure}
\begin{center}
\includegraphics[angle=0.,scale=.9]{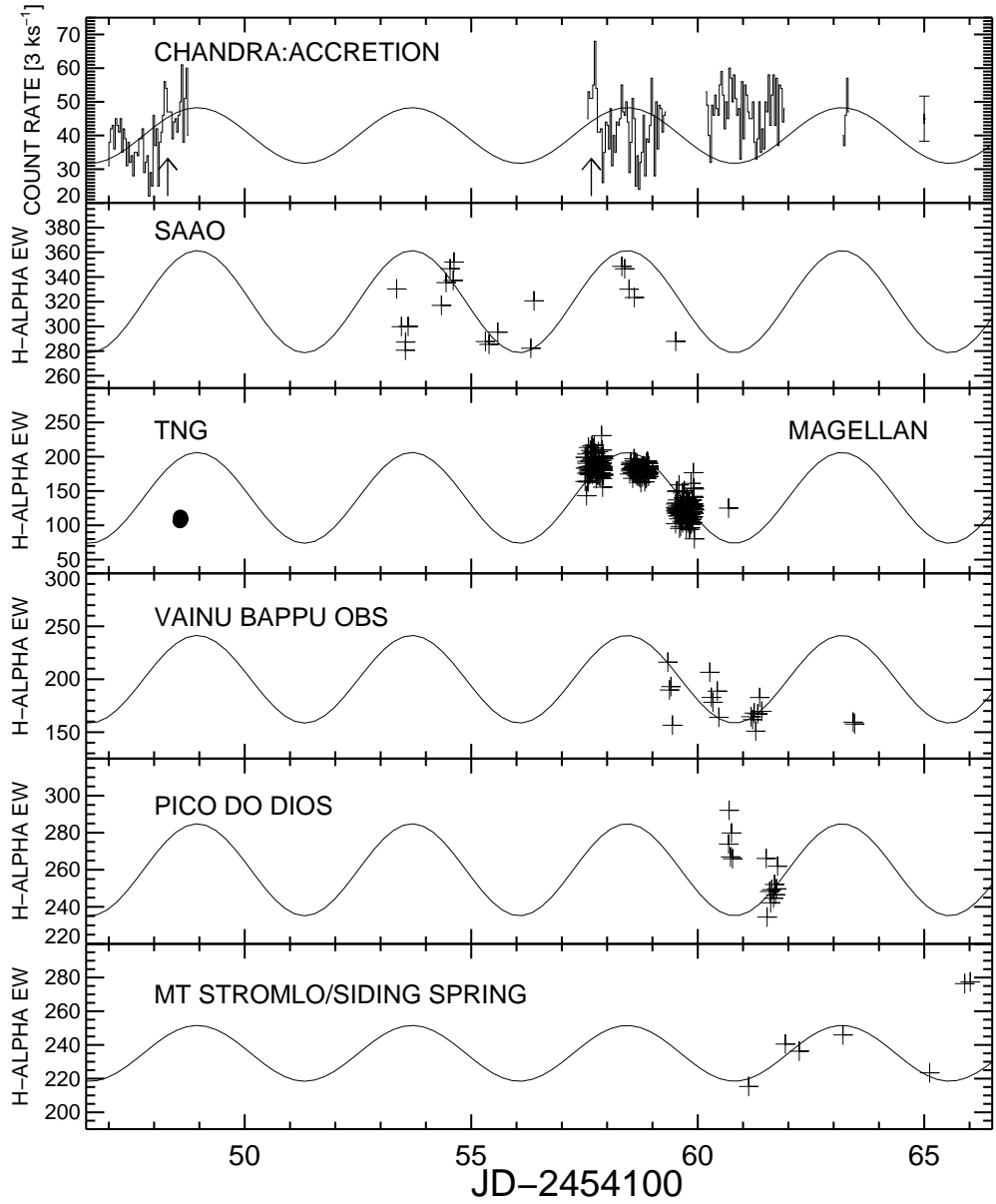}
\caption{\chandra\ accretion line flux binned to 3 ks and H-\gal\ equivalent widths for all of
  the spectroscopic observations.  A sine curve with the period of 4.744 d from
  WASP-S photometry and arbitrary amplitude is overlaid, and placed to guide the eye as to
the H-\gal\ strength.  }
\vspace{-0.5in}
\end{center}
\end{figure}
\clearpage 
\begin{figure}
\plottwo{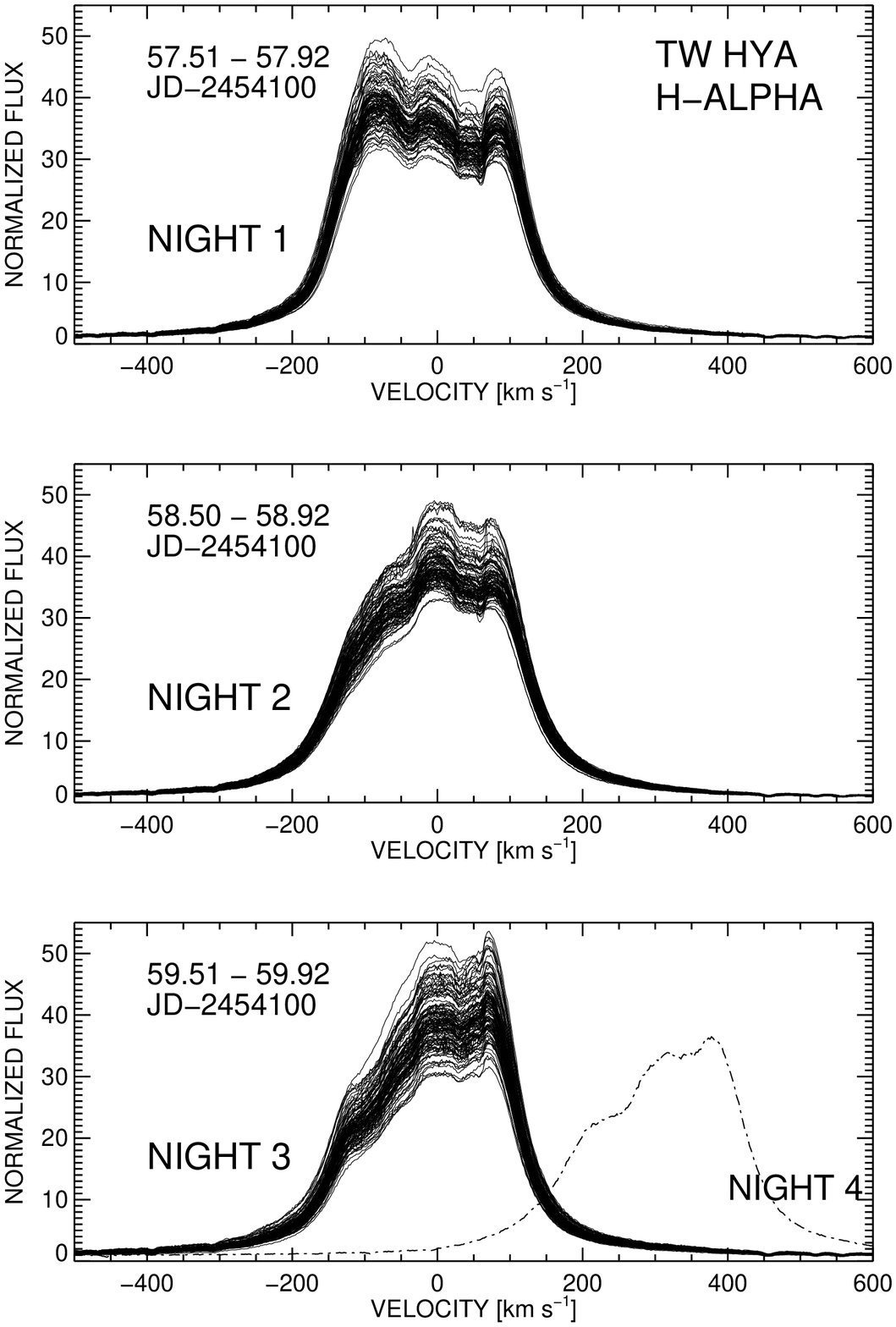}{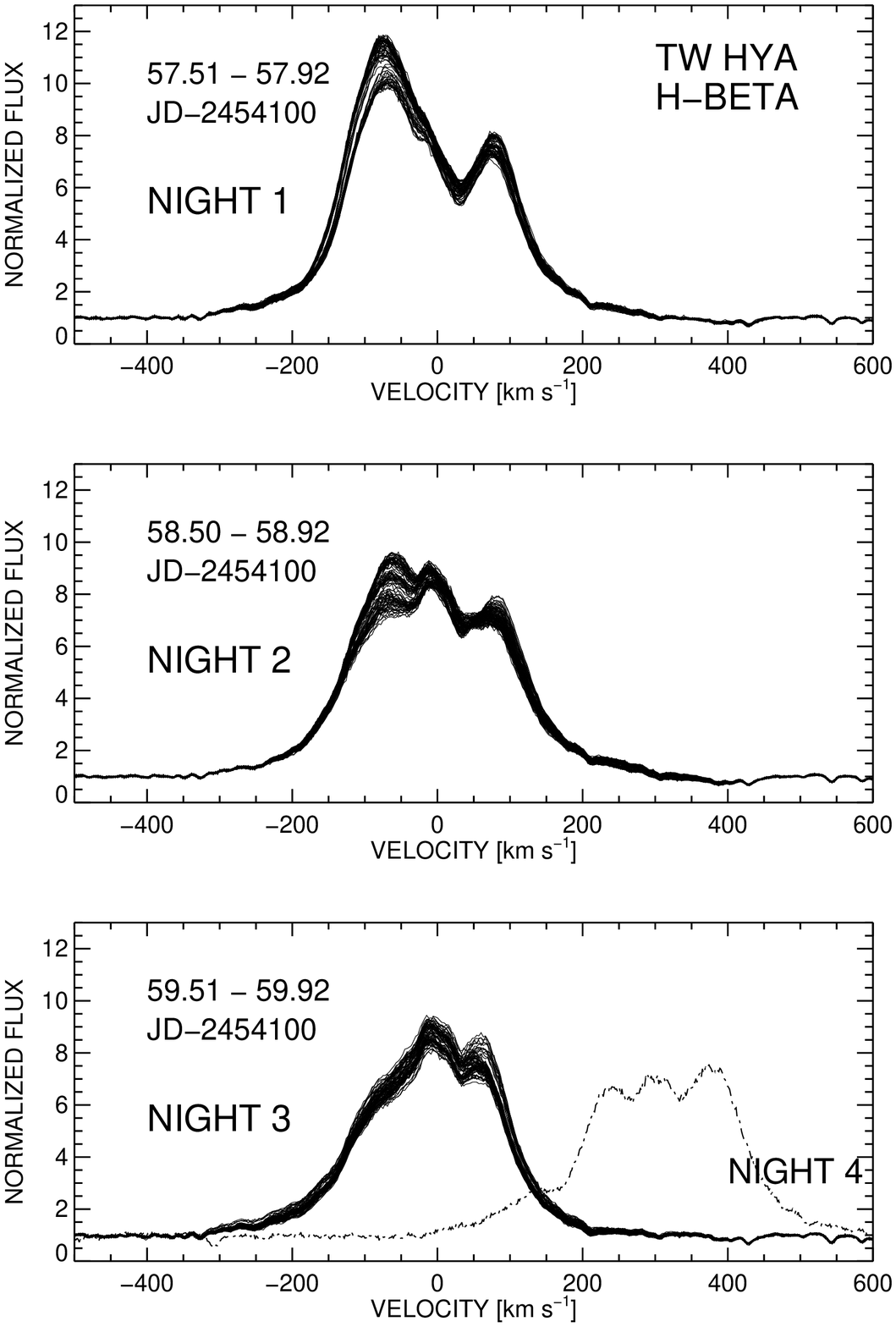}
\caption{Four consecutive nights of continuous observations of the
  H-\gal\ ({\it left panel}) and H-$\beta$ ({\it right panel})
  profiles from Magellan reveal substantial systematic changes principally on the
short-wavelength side of the emission profile.  The profile from night
  4 is offset by 300~\kms\ for display.}
\end{figure}
\clearpage

\begin{figure}
\includegraphics[angle=90.,scale=0.7]{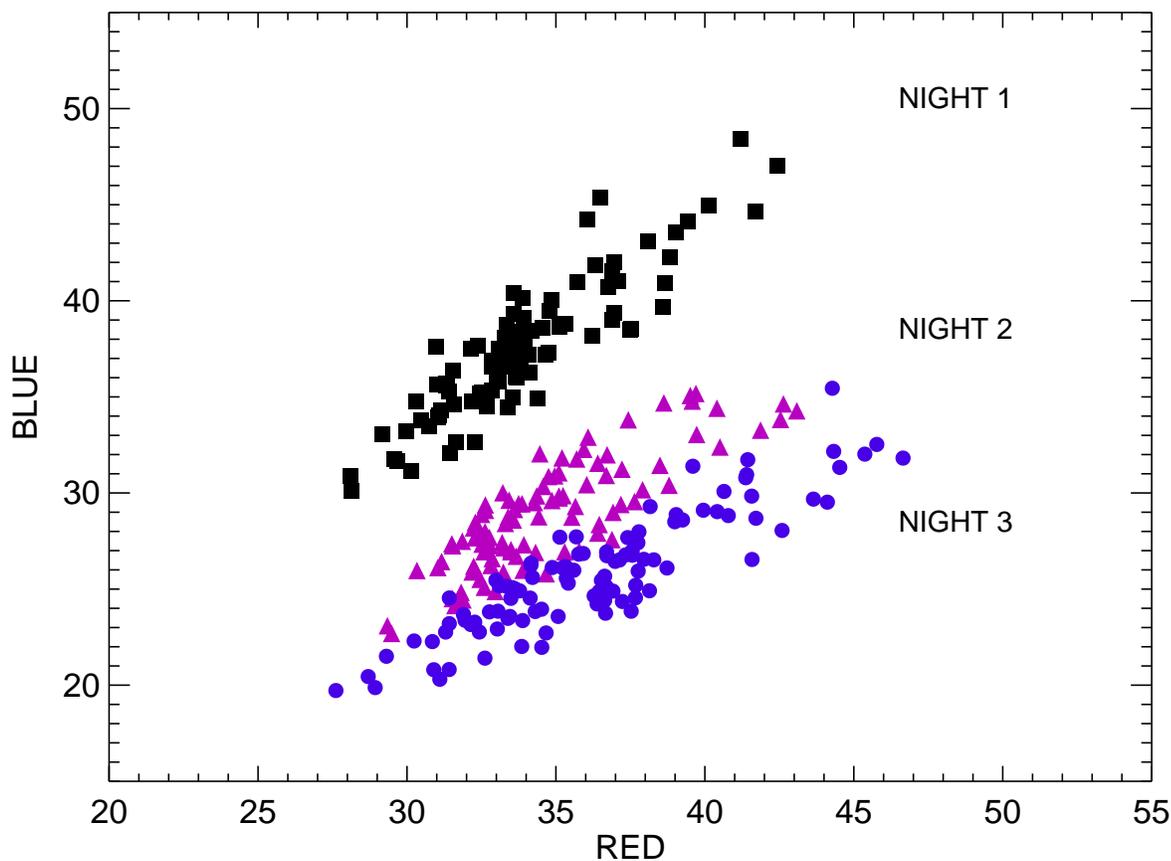}

\caption{Strength of the blue wing in the H-$\alpha$ emission line as a function
of the strength of the red wing from three sequential
nights of Magellan spectra: Day 57-59. Both sides of the line are
correlated indicating that a single process produces this intrinsically broad
line.  Filled
squares ($\blacksquare$) correspond to Night 1 (starting $\sim$JD 2454157.5 ); 
filled triangles ($\blacktriangle$ )
correspond to Night 2 (JD 2454158.5 ); filled circles ($\bullet$ ) mark Night 3
(JD 2454159.5). Fluxes are measured in a 1\AA\ wide band positioned
$\pm$2\AA\ from line center in continuum-normalized spectra.  The decreasing slope over the 3 nights
results from the increasing wind opacity that weakens the 'blue' side
of the line profile. }
\end{figure}

\clearpage

\begin{figure}
\includegraphics[angle=90.,scale=0.7]{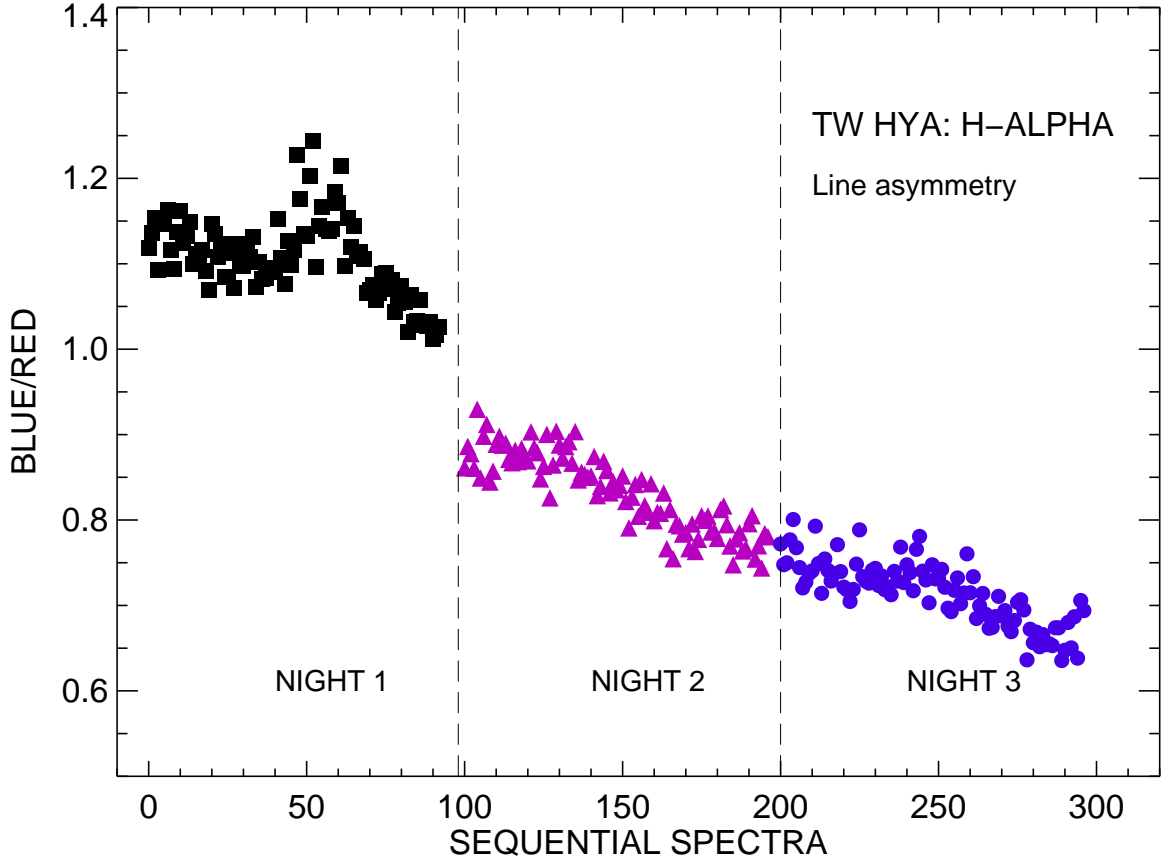}
\caption{Blue:red asymmetry as a function of time over 3 consecutive
nights. The 'blue' and 'red' fluxes are taken from a 1\AA\ wide band
positioned $\pm$2\AA\ from line center in continuum-normalized spectra.
A systematic decrease of the ratio occurs over 3 successive
nights following generally higher activity during the first night
after the accretion event (corresponding to Spectrum $\sim$50) described in Section 5.2.}
\end{figure}

\clearpage
\begin{figure}
\begin{center}
\includegraphics[angle=0.,scale=0.7]{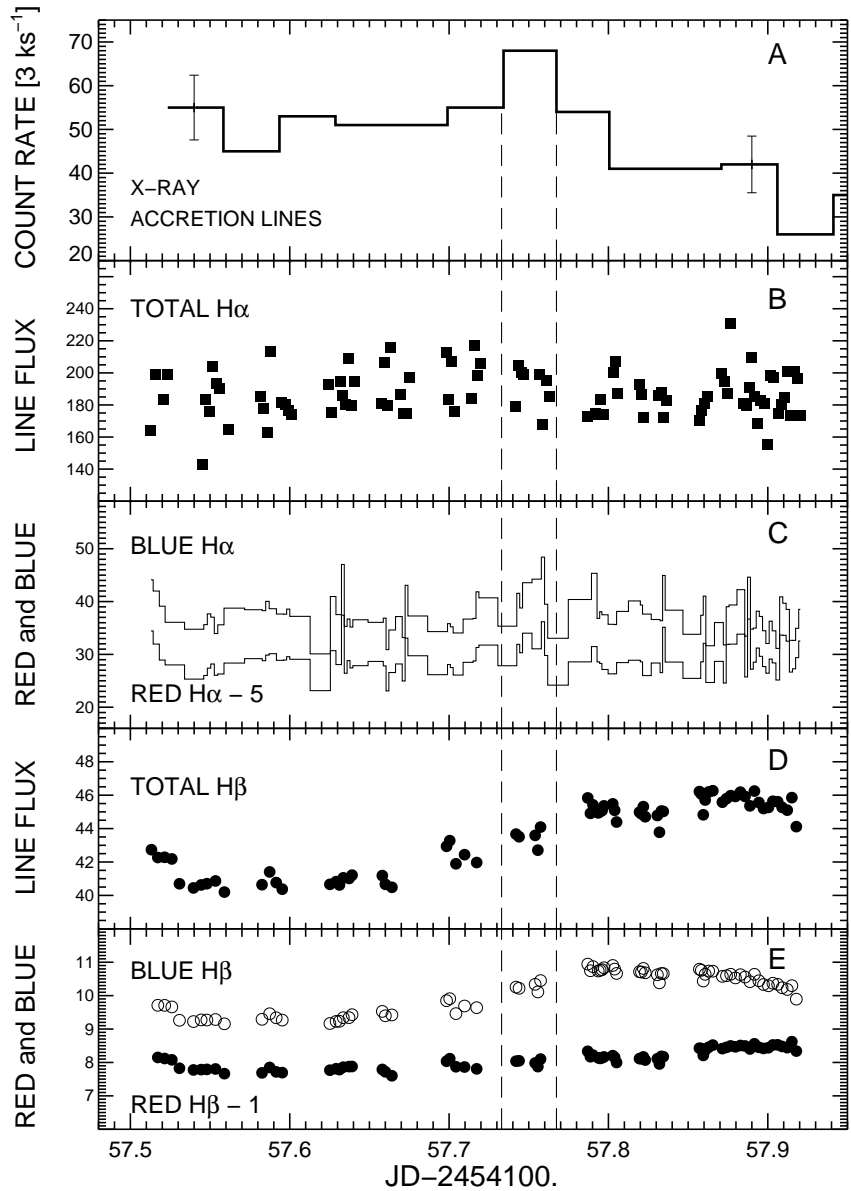}
\caption{The H-\gal\ and H-\gb\ flux during the accretion event of Day
  57.74. Also included are the fluxes from a 1\AA\ region on the short (blue) and long (red)
  wavelength sides of each line.  The red H-$\alpha$ flux has been lowered by 5
  units in this figure; similarly the red H-$\beta$ flux has been lowered
by 1 unit for display.}

\end{center}
\end{figure}

\clearpage
\begin{figure}
\begin{center}

\includegraphics[angle=0.,scale=0.6]{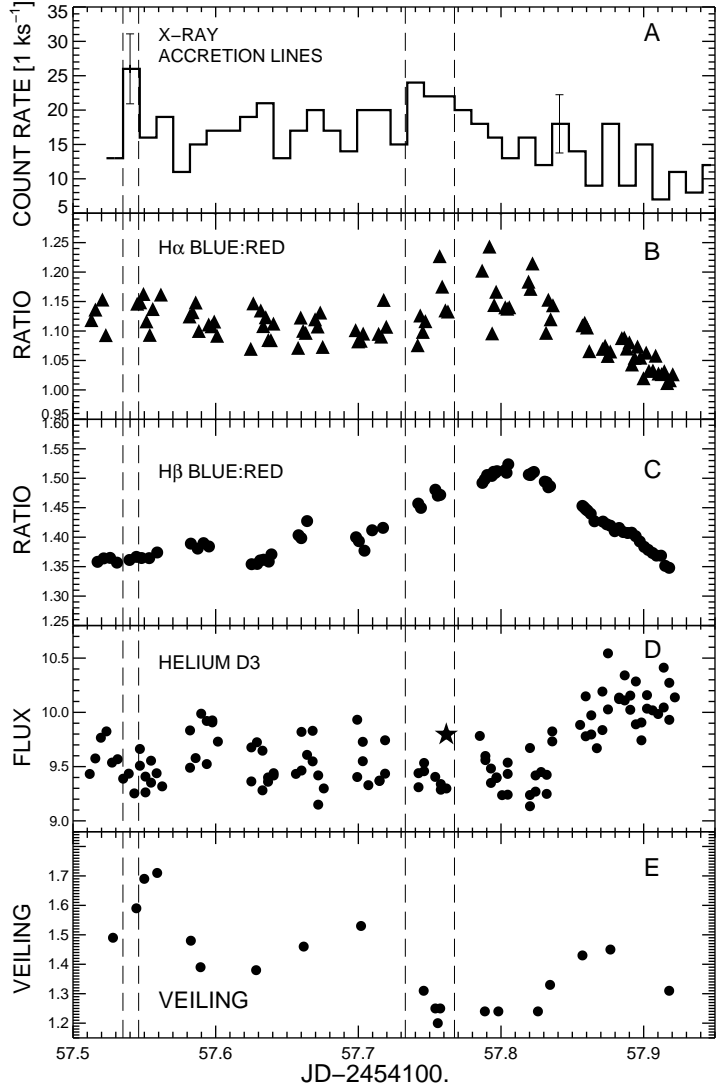}
\caption{Changes in the profile characteristics of the H-\gal\ and
  \ion{He}{1} (D3, 5876\AA) lines
during an `accretion event' as indicated by the increase in flux of X-ray
line emission (binned at 1ks)  formed in the accretion shock ({\it
  Panel A}). 
{\it Panel B:}  The total flux
of H-\gal\ normalized to the continuum does not 
change (see Figure 11), but the flux ratio of a 1\AA\ band on the
short (blue) and long (red) wavelengths of the H-\gal\ emission displays
an increased asymmetry 
near Day 57.75.  {\it Panel C}: Flux ratio of the H-\gb\ line, short
  wavelength to long wavelength measured in similar fashion as H-\gal.   
{\it Panel D:} The continuum-normalized flux of the \ion{He}{1} D3 line (\gla5876). The star
  symbol marks the spectrum taken at 06:18 (Day 57.763) where
the wings of the helium line have broadened (see Figure 14).  Veiling 
  ({\it Panel E})  is also shown (described in Section  9). }
\end{center}
\end{figure}
\clearpage

\begin{figure}
\begin{center}
\includegraphics[angle=90.,
  scale=0.6]{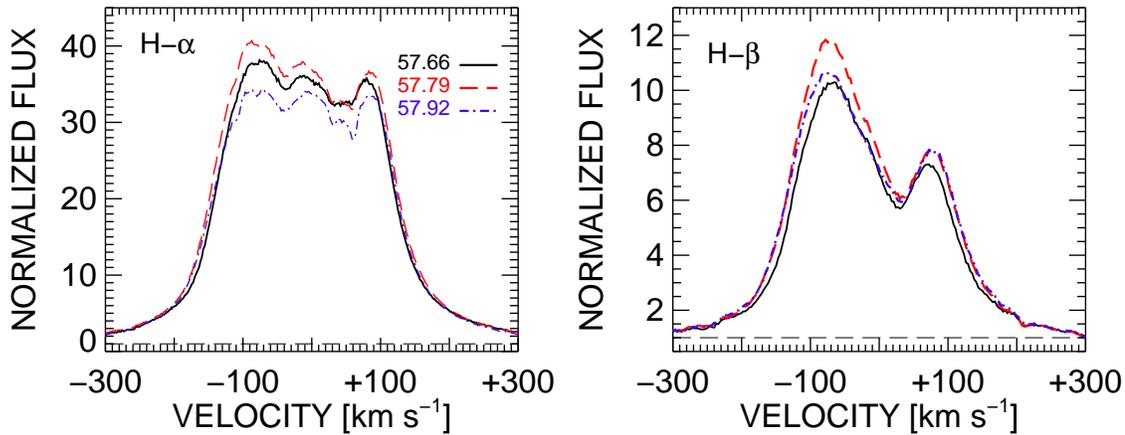}
\caption{Profiles from Magellan spectra that
sample the time (Day 57.66)   before the long X-ray accretion event, the onset of
H-\gal\ asymmetry (Day 57.79) noted in Figure 12, and at the end
of the observing night (Day 57.92).   {\it Left
    panel:} H-\gal: The total line flux increases slightly between
57.66 and 57.79 in these spectra, but Figure 11 shows this increase is not
significant. At the end of the observing night (marked 57.92), the
H-\gal\ line flux is less and the profile more symmetric. {\it Right
    panel:} H-\gb\ profile showing a similar flux change and
the signature of constant inflow. Both emission lines show
increased blue asymmetry indicating increased downflow directly after the
X-ray accretion event, but this blue asymmetry decreases starting at
Day 57.8, about 1.5 hours after the X-ray accretion event.}
\end{center}
\end{figure}
 
\clearpage
\begin{figure}
\begin{center}
\includegraphics[angle=90,scale=0.8]{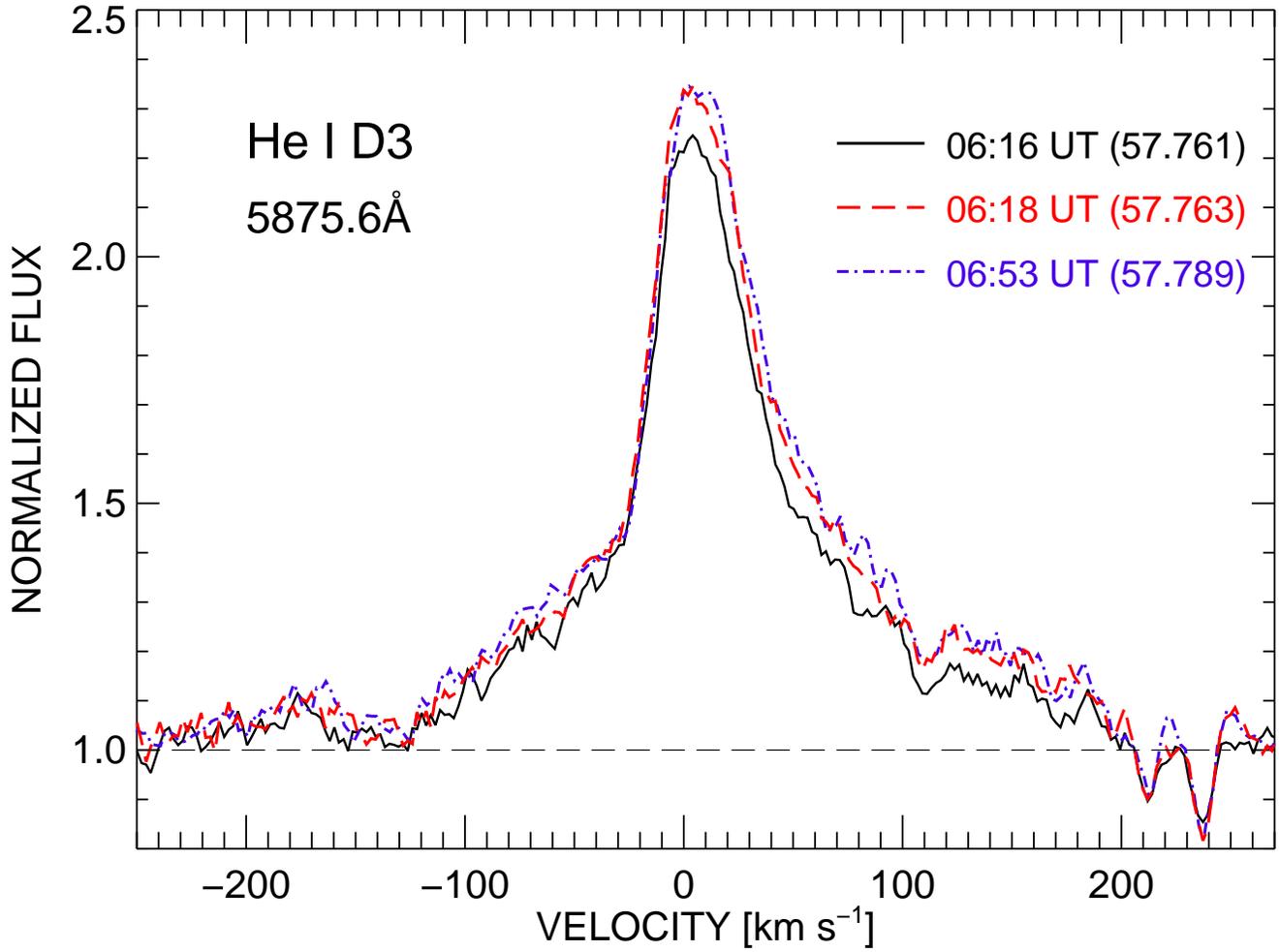}
\caption{The \ion{He}{1} D3 line directly following the change in the
  H-\gal\ line asymmetry.  The local UT time (26 February 2007) of
  mid-exposure is
noted for the three spectra, and the mid-exposure time of the spectrum 
obtained at 06:18 UT (Day 57.763)  is marked by a star in Figure 12.}

\end{center}
\end{figure}

\begin{figure}
\begin{center}

\includegraphics[angle=0,scale=0.7]{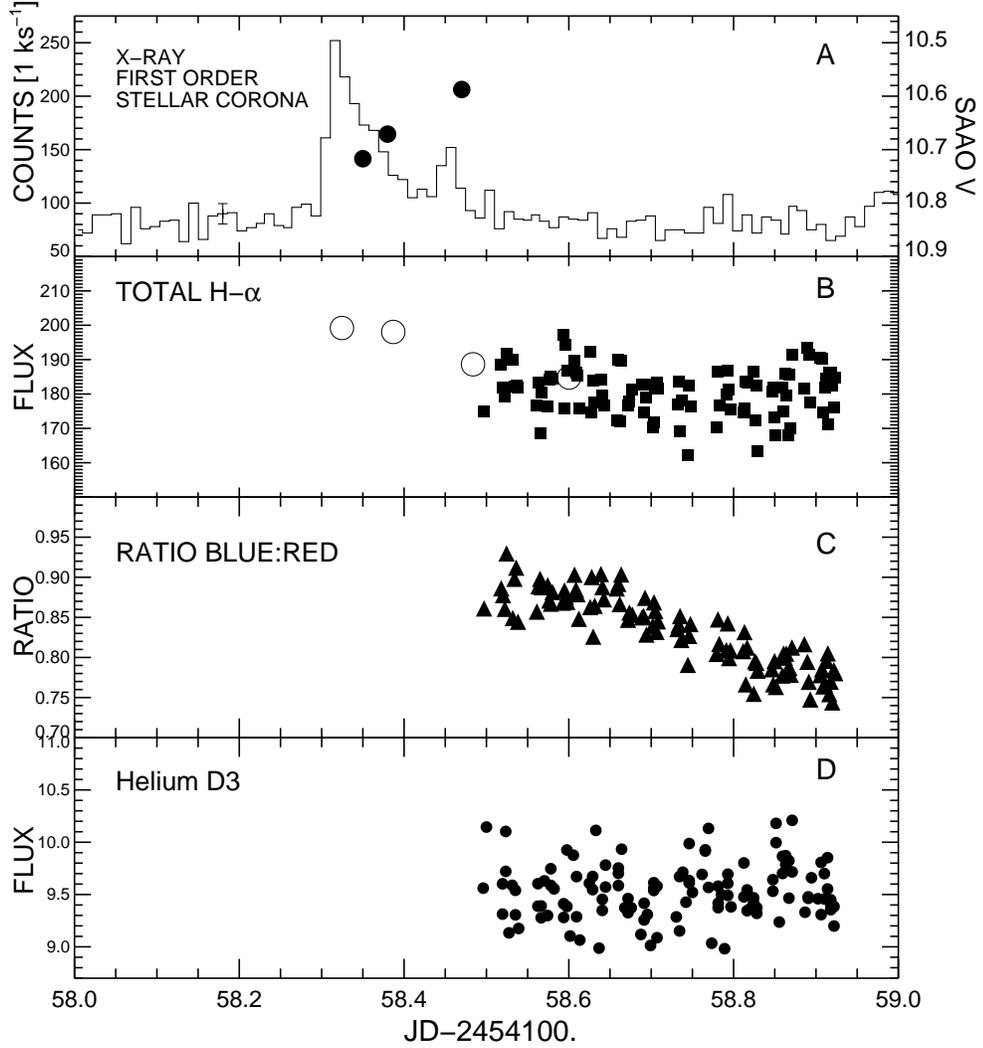}

\caption{H-\gal\ characteristics near the coronal flaring event. {\it
    Panel A:} Total X-ray flux (corona) binned in 1 ks intervals at
    the time of the coronal flare.  The filled circles represent the
    simultaneous SAAO photometry with the scale on the right side
    axis. {\it Panel B:} The total flux in the H-$\alpha$ line as
    measured in the spectra from the SAAO ($\bigcirc$) and spectra
    taken at Magellan ($\blacksquare$). {\it Panel C:} Ratio of
    the strength of the blue wing to the red wing of the H-$\alpha$
    line measured from the Magellan spectra. {\it Panel D:} Flux of
    the \ion{He}{1} emission line at \gla 5876. These line
    characteristics differ from those found following the accretion
    event shown in Figure 12; a distinct accretion event is not
    seen in the X-ray data at this time (see Figure 7).}
\end{center}
\end{figure}
\clearpage

\begin{figure}
\includegraphics[angle=0.,scale=0.8]{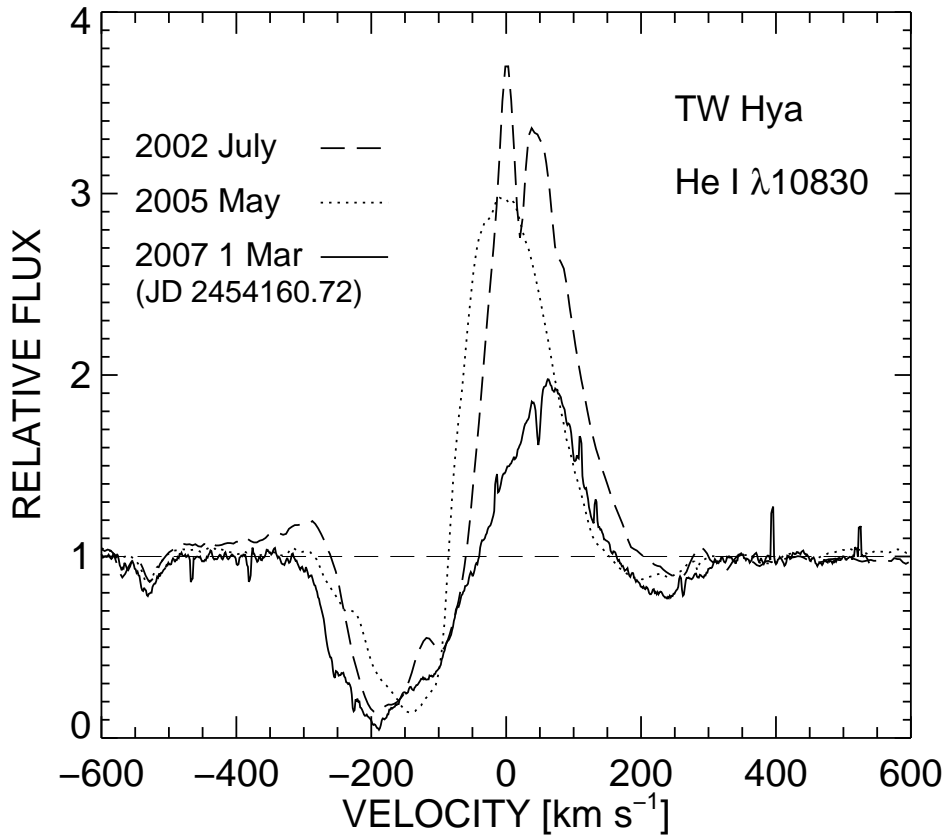}
\caption{PHOENIX spectrum of He I, 10830\AA\ obtained at Gemini-S on 
on 1 March 2007 (JD 2454160.5), as compared
to spectra obtained previously with KECK:NIRSPEC at slightly lower resolution (Dupree
et al. 2005). The emission in 2007 is substantially weaker, and wind 
absorption extends to higher velocities than in previous observations.
Stronger absorption occurs  
below the continuum on the long wavelength side of the line.  The
total extent of the line appears to be $\pm$325 \kms.} 
\end{figure}
\clearpage
\begin{figure}
\begin{center}
\includegraphics[angle=0.,scale=0.8]{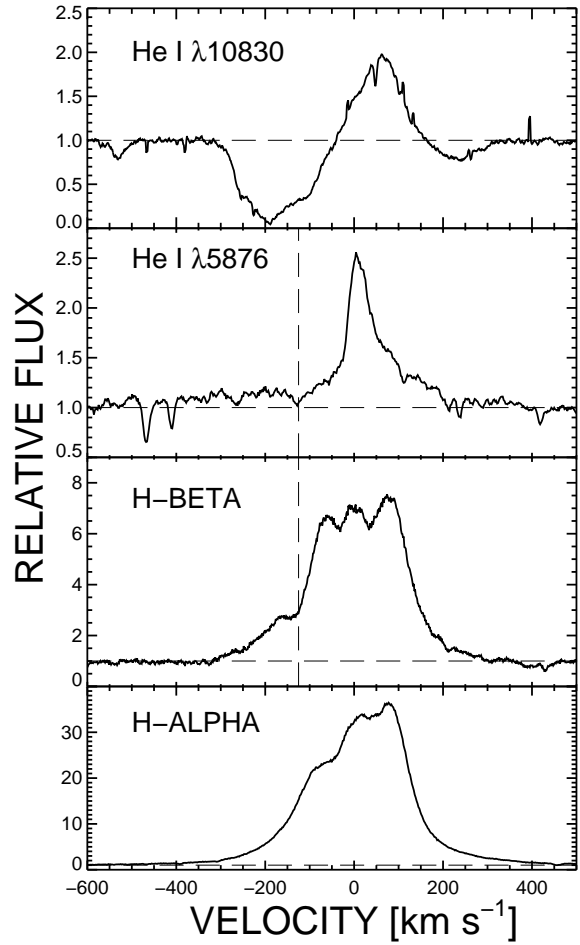}
\vspace*{0.3 in}
\caption{Three major diagnostic emission lines observed within an hour
of the \ion{He}{1} near-infrared 10830\AA\ transition on Day
60. Evidence
of an outflowing wind is apparent in all lines with outflow velocities
increasing from H-\gb\ and the 5876\AA\ line to the 10830\AA\
transition. Inflow can also be noted in the subcontinuum absorption
of the 10830\AA\ line.}
\end{center}
\end{figure}

\begin{figure}
\begin{center}
\includegraphics[angle=0,scale=1.0]{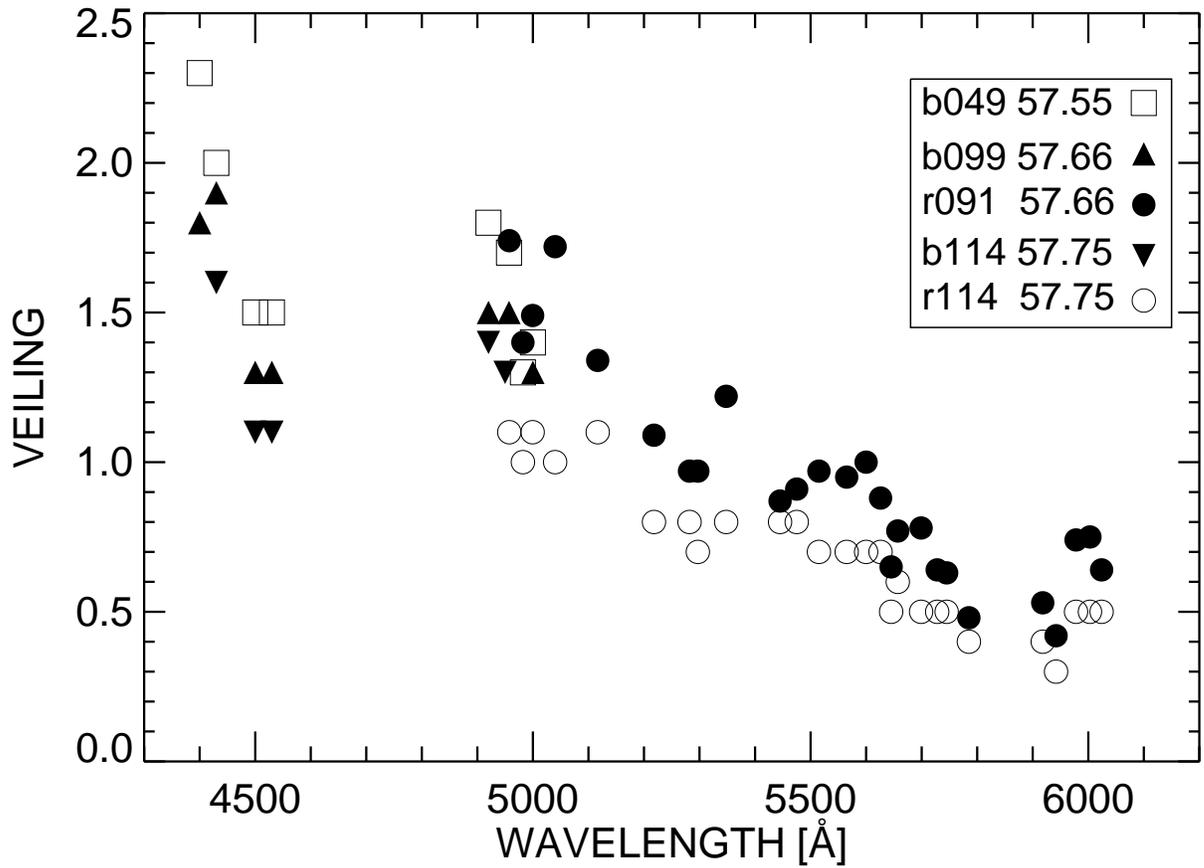}
\caption{Veiling as a function of wavelength during the night of
26 February 2007 (UT). The spectrum [either blue (b) or red (r) side of MIKE,
and the Day (JD-245100) are noted]. The veiling  increases
towards shorter wavelengths and  the value of the  veiling decreases 
during the first part of the night.  The 'gap' between 4600\AA\ and 4800\AA\  occurs
because photospheric lines in this region are shallow, and it is
difficult to determine a value of the veiling.}
\end{center} 
\end{figure}
\clearpage
\begin{figure}
\begin{center}
\includegraphics[angle=0,scale=0.7]{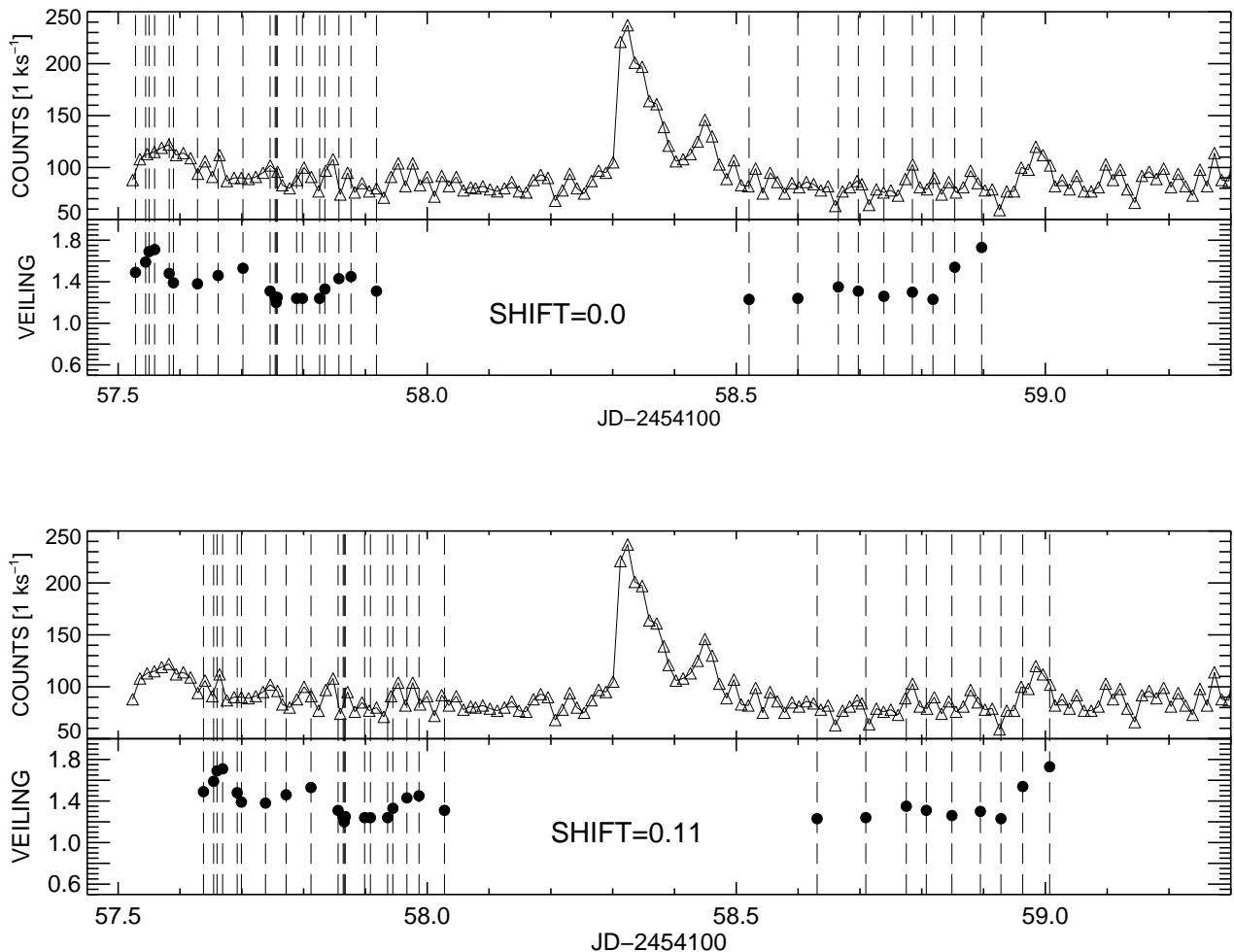}
\caption{Total (coronal) X-ray flux and average blue veiling with no time
  shift ({\it top 2 panels}) and a time shift of 0.11 d applied to the
veiling measures ({\it lower 2 panels}), suggesting that the
  appearance of a hot
photospheric region (indicated by veiling) precedes the increase in
  coronal X-ray flux. Error bars (1$\sigma$) for the X-ray flux
  measures are smaller than the symbol size.}
\end{center}
\end{figure}
\clearpage

\begin{figure}
\begin{center}
\includegraphics[angle=0.,scale=0.7]{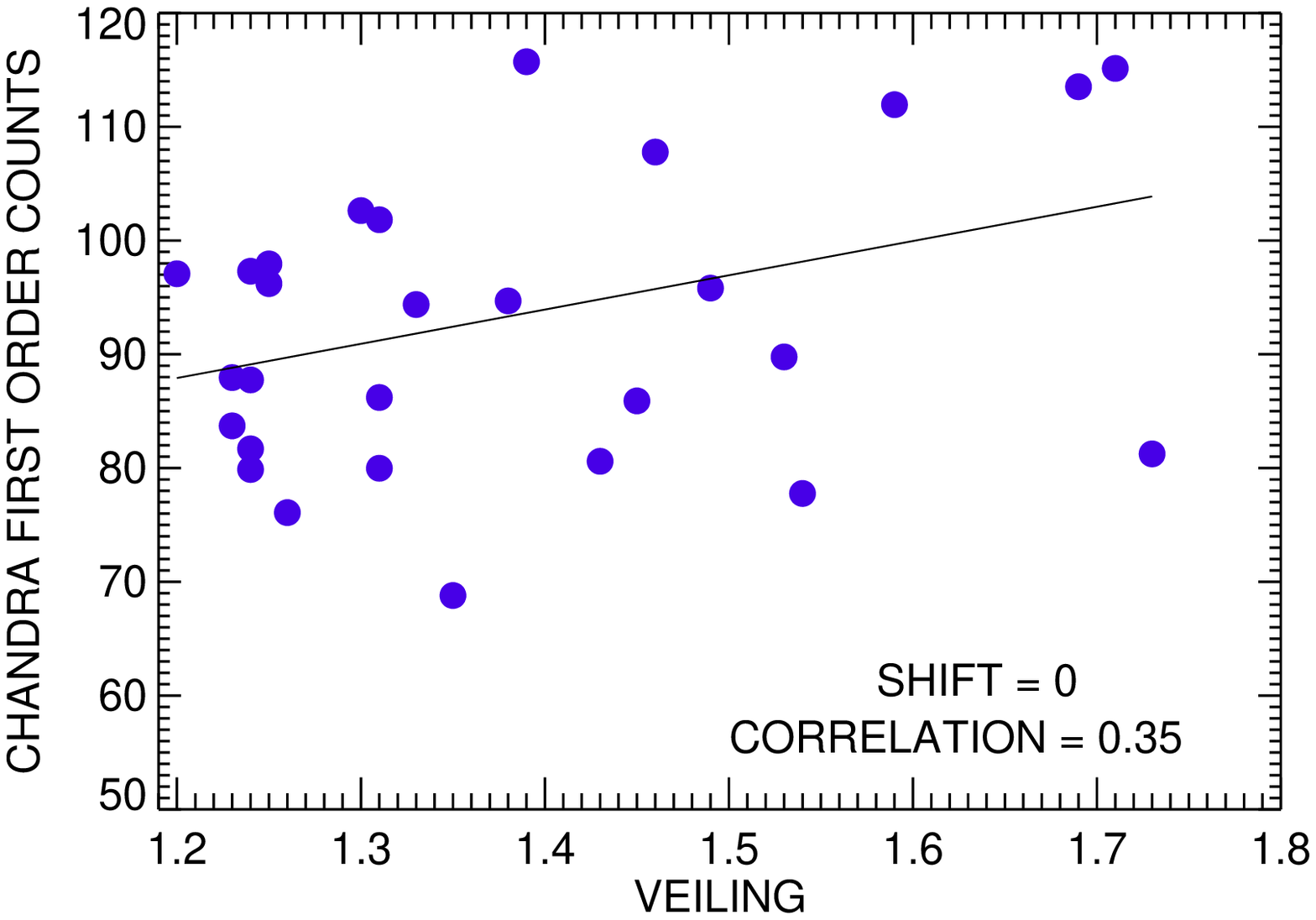}
\includegraphics[angle=0.,scale=0.7]{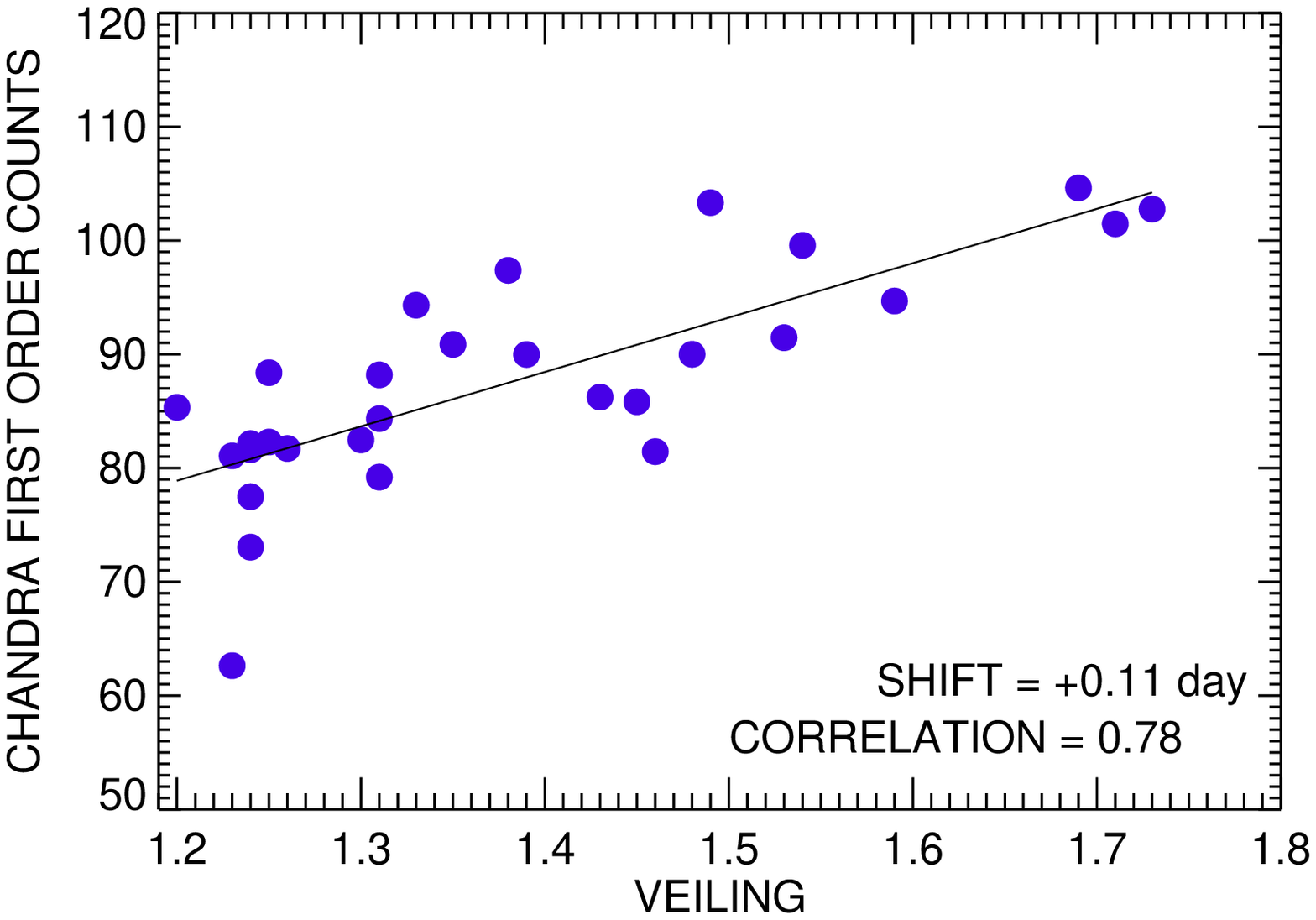}
\caption{Total first order \chandra\ counts interpolated to the
times of the spectra as a function of
the average blue veiling (4500--5000\AA). {\it Top panel:} No time
shift between veiling measure and coronal X-ray flux. {\it Lower
  panel:} The veiling measure is shifted by +0.11 day where the
agreement
between the coronal flux and veiling is substantially improved. This
suggests that the corona responds 0.11 d ($\sim$ 2.6 hr) later to an increase in the
veiling.}
\end{center}
\end{figure}
\clearpage


\begin{figure}
\begin{center}
\includegraphics[angle=0.,scale=0.7]{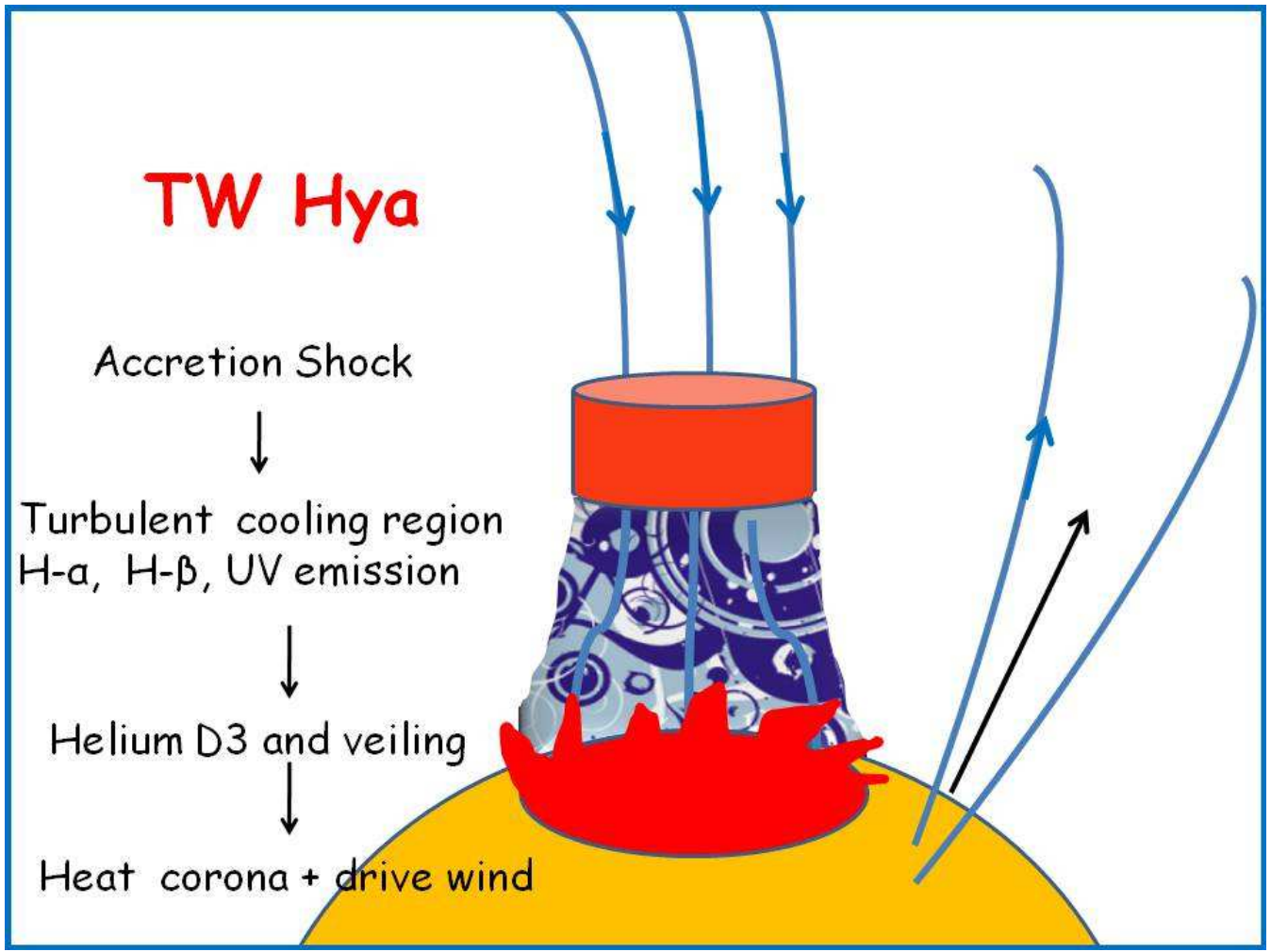}
\vspace*{1 in}
\caption{Schematic cartoon suggesting the sequential steps in the
  accretion process.}

\end{center}
\end{figure}
\clearpage

\appendix
\section{Effect of Water Vapor} 

Water vapor absorption in the Earth's atmosphere occurs
at wavelengths overlapping the H-$\alpha$ profile, but only
in severe instances does it affect the total line flux.  
Examples of absorption in the emission profile are shown in
the accompanying figure. Wavelengths are taken from the HItran
database (Rothman \etal\ 2009).  The strongest line in 
this region estimated from the R. Kurucz synthesis of the
solar spectrum occurs at 6564.195\AA.  It is clear that the
peak in the H-$\alpha$ profile at longest wavelengths can
be affected by water vapor absorption, although the lines
are narrow.  The 3 profiles
in Figure 21 show that the wind absorption is   
dominant in shaping the overall profile.  The spectrum
at the highest airmass of 3.3 was taken under adverse
conditions during the monsoon season in India.


\begin{figure}
\begin{center}

\includegraphics[angle=90.,scale=0.7]{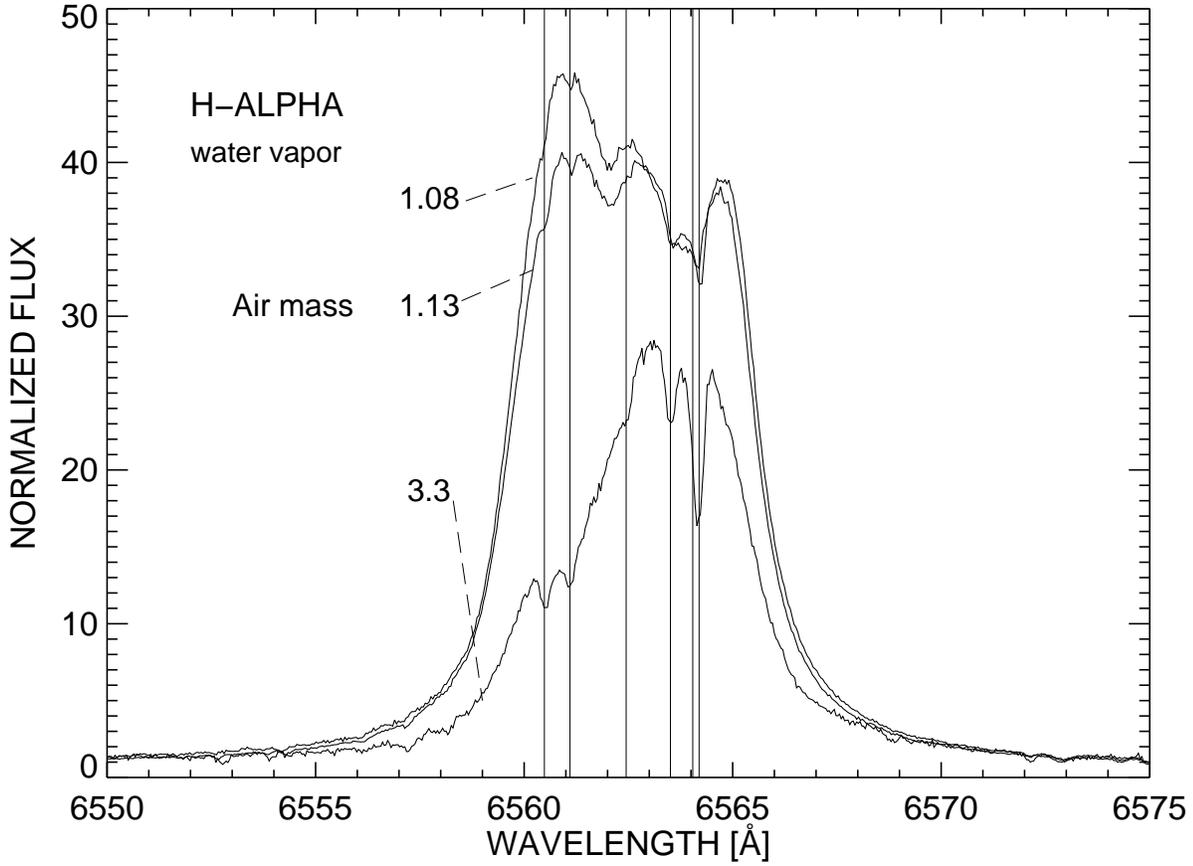}
\caption{Three profiles of H-$\alpha$ to illustrate the presence of
  water vapor in the profile.  Spectra were taken at 3 different air
  masses and are aligned on the rest wavelengths of the water vapor as
  determined in the  HITRAN database (Rothman \etal\ 2009).}
\end{center}
\end{figure}

\clearpage
\begin{deluxetable}{llr}
\def\a{\phantom{0}}
\def\b{\phantom{00}}
\tablecolumns{3}
\tablewidth{0pt}
\tablenum{1}
\tablecaption{CHANDRA ACIS-S/HETG Observations}
\tablehead{
\colhead{OBSID}&
\colhead{JD Start} &
\colhead{Exposure}\\
\colhead{}&
\colhead{($-$2454100)}&
\colhead{(ks)}\\
}
\startdata
7435 & 46.986&152 \\
7437 & 57.519&160 \\
7436 & 60.146&162 \\ 
7438 & 63.156 &23 \\
\enddata
\end{deluxetable}

\begin{deluxetable}{llllccrc}
\def\a{\phantom{0}}
\def\b{\phantom{00}}
\tablecolumns{6}
\tablewidth{0pt}\tablenum{2}
\tablecaption{Ground-Based Observations}
\tablehead{
\colhead{Observatory}   &     
\colhead{Instrument}  &     
\colhead{Data}  &
\colhead{Observer} &
 \colhead{JD Start } &
\colhead{JD End} \\
\colhead{Telescope}&
\colhead{} &
\colhead{} &
\colhead{} &
\colhead{($-$2454100)} &
\colhead{($-$2454100)} 
}
\startdata
WASP-S& & Photometry & Robotic & 48.33& 67.48   \\
TNG &SARG&Spectra & DDT& 48.57 & 49.60 \\
SAAO & Cassegrain &  Spectra &Crause/Lawson &53.35 & 59.53 \\
SAAO&         &Photometry&Crause/Lawson&53.56 & 59.53\\
ASAS&         &Photometry&Robotic&47.69 & 63.72\\
VBO & Echelle & Spectra & Mallik &59.25 & 63.48 \\
Magellan/Clay & MIKE & Spectra & Dupree & 57.51 & 59.93 \\
Magellan/Clay & MIKE & Spectra & Bonanos & 60.68 & 60.69\\
Pico dos Dias & Coud\'e & Spectra & Luna & 60.68 & 61.77\\
Gemini-S & PHOENIX & Spectra & Schuler& 60.71 & 60.74 \\ 
SSO & Echelle & Spectra & Bessell/Lawson & 61.12 & 66.02 \\
\enddata
\end{deluxetable}


\begin{deluxetable}{llcllc}
\def\a{\phantom{0}}
\def\b{\phantom{00}}
\tablecolumns{6}
\tablewidth{0pt}\tablenum{3}
\tablecaption{Blue Veiling Parameter}
\tablehead{
\colhead{DAY}&
\colhead{Veiling} &
\colhead{Spectrum}&
\colhead{DAY}&
\colhead{Veiling} &
\colhead{Spectrum}\\
\colhead{(JD$-$2454100)}&
\colhead{}&
\colhead{}&
\colhead{(JD$-$2454100)}&
\colhead{}&
\colhead{}
}
\startdata
57.5280 &    1.49   &     b044  &  57.8259  &    1.24    &    b138 \\ 
57.5444  &   1.59    &    b047  & 57.8344   &    1.33    &     b143 \\
57.5500  &   1.69   &     b049  &  57.8571  &    1.43  &     b148 \\ 
57.5590  &    1.71   &     b050 &  57.8767  &    1.45  &     b157 \\   
57.5826  &   1.48    &    b069  & 57.9179   &   1.31   &    b170 \\
57.5894  &    1.39  &      b071 & 58.5207  &     1.23  &     b050 \\ 
57.6284  &   1.38   &     b090  & 58.5995  &     1.24  &     b068 \\
57.6617  &   1.46   &     b099  & 58.6650  &    1.35   &    b091 \\ 
57.7019   &   1.53   &     b106 & 58.6974  &    1.31   &    b102 \\
57.7458  &   1.31   &     b114  & 58.7388  &    1.26  &     b114 \\
57.7540   &   1.25   &     b115 & 58.7848  &    1.30   &    b131 \\ 
57.7557  &    1.20   &     b116 & 58.8181  &    1.23  &     b143 \\ 
57.7574  &    1.25    &    b117 &58.8534   &    1.54  &     b166 \\ 
57.7887   &   1.24    &    b121 & 58.8969  &    1.73   &    b181  \\    
57.7982  &   1.24    &    b127 &           &           &          \\

\enddata
\end{deluxetable}  


\end{document}